\documentclass[11pt]{article}
\usepackage{amsthm,amsmath,latexsym,multibox,amssymb,amsfonts,amsbsy}
\usepackage{graphicx,lscape,fancyhdr,array,stmaryrd,euscript,wrapfig}
\usepackage{cite,wrapfig,color}

\renewcommand{\theequation}{\arabic{section}.\arabic{equation}}
\newcolumntype{x}[1]{%
>{\centering\hspace{0pt}}m{#1}}%
\newcolumntype{w}[1]{%
>{\raggedright\hspace{0pt}}m{#1}}%
\newcolumntype{z}[1]{%
>{\raggedleft\hspace{0pt}}m{#1}}%

\newcommand{\dSAdS}{{\ensuremath{(A)dS_d}\,}}
\newcommand{\AdS}{{\ensuremath{AdS_d}\,}}

\newcommand{\ads}{{\ensuremath{\mathfrak{so}(d-1,2)}}}

\newcommand{\iso}{{\ensuremath{\mathfrak{iso}(d-1,1)}}}
\newcommand{\lorentz}{{\ensuremath{\mathfrak{so}(d-1,1)}}}

\newcommand{\msv}{{\ensuremath{\mathfrak{so}(d-1)}}}
\newcommand{\mls}{{\ensuremath{\mathfrak{so}(d-2)}}}
\newcommand{\pl}{\partial}
\newcommand{\fm}[1]{_{\mathbf{{#1}}}}
\newcommand{\be}{\begin{equation}}
\newcommand{\ee}{\end{equation}}

\newcommand{\Ya}[1]{{\ensuremath{[#1]}}}

\newcommand{\ba}{{\ensuremath{\mathbf{a}}}}
\newcommand{\bb}{{\ensuremath{\mathbf{b}}}}

\newcommand{\aA}{{\ensuremath{\mathcal{A}}}}
\newcommand{\aB}{{\ensuremath{\mathcal{B}}}}
\newcommand{\aC}{{\ensuremath{\mathcal{C}}}}
\newcommand{\aD}{{\ensuremath{\mathcal{D}}}}

\newcommand{\DL}{{D}}

\newcommand{\DO}{{D_{0}}}

\newcommand{\fud}[2]{{^{#1}_{\phantom{#1}#2}}}

\newcommand{\fudu}[3]{{^{#1\phantom{#2}#3}_{\phantom{#1}#2}}}

\newcommand{\smallpic}[1]{{\unitlength=0.2mm#1}}

\definecolor{rougef}{rgb}{0.56,0,0}
\definecolor{vertf}{rgb}{0,0.5,0}
\definecolor{bleuf}{rgb}{0,0,0.8}

\newcommand{\bep}{\begin{picture}}
\newcommand{\eep}{\end{picture}}

\newcounter{YoungHeight}\newcounter{YoungWidth}

\newcounter{Mul1}\newcounter{Mul2}\newcounter{Mul3}\newcounter{Mul4}
\newcounter{A1}\newcounter{A2}
\newcounter{B3}

\newcommand{\Add}[3]{\setcounter{#1}{#2}\addtocounter{#1}{#3}}

\newcommand{\Length}[1]{#10}
\newcommand{\YoungScale}{}

\newcommand{\BlockA}[2]{{\YoungScale\bep(\Length{#1},\Length{#2}){\Add{A1}{#1}{1}\Add{A2}{#2}{1}}%
\multiput(0,0)(10,0){\value{A1}}{\line(0,1){\Length{#2}}}\multiput(0,0)(0,10){\value{A2}}{\line(1,0){\Length{#1}}}%
\setcounter{YoungHeight}{\Length{#2}}\setcounter{YoungWidth}{\Length{#1}}\eep}}

\newcommand{\BlockB}[4]{{\YoungScale\Add{B3}{\Length{#2}}{\Length{#4}}%
\bep(\Length{#1},\value{B3})\put(0,\Length{#4}){\BlockA{#1}{#2}}%
\put(0,0){\BlockA{#3}{#4}}\setcounter{YoungHeight}{\value{B3}}\setcounter{YoungWidth}{\Length{#1}}\eep}}

\newcommand{\YoungBA}{\BlockB{2}{1}{1}{1}}%

\begin{document}
\renewcommand{\thefootnote}{\fnsymbol{footnote}}
\begin{flushright}
\vspace{1mm}
\end{flushright}

\vspace{1cm}

\begin{center}
{\bf \Large  Gravitational cubic interactions for a simple
mixed-symmetry gauge field in AdS and flat backgrounds}

\vspace{2cm}

\textsc{Nicolas Boulanger\footnote{Research Associate of the Fund
for Scientific Research-FNRS (Belgium);
nicolas.boulanger@umons.ac.be}, E.D.
Skvortsov\footnote{skvortsov@lpi.ru} and Yu. M.
Zinoviev\footnote{Yurii.Zinoviev@ihep.ru}}

\vspace{2cm}

{\em${}^*$ Service de M\'ecanique et Gravitation, Universit\'e de Mons -- UMONS\\
20 Place du Parc, 7000 Mons (Belgium)} \vspace*{.2cm}

{\em${}^\dag$ P.N.Lebedev Physical Institute, Leninsky prospect 53,
119991, Moscow (Russia)} \vspace*{.2cm}

{\em${}^\ddag$ Institute for High Energy Physics Protvino, Moscow
Region, 142280, Russia} \vspace{1cm}

\end{center}

\vspace{0.5cm}
\begin{abstract}
Cubic interactions between the simplest mixed-symmetry gauge field
and gravity are constructed in anti-de Sitter (AdS) and flat backgrounds.
Nonabelian cubic interactions are obtained in AdS following various
perturbative methods including the Fradkin--Vasiliev construction,
with and without St\"uckelberg fields. The action that features
the maximal number of St\"uckelberg fields can be considered in the flat limit
without loss of physical degrees of freedom. The resulting interactions in
flat space are compared with a classification of vertices obtained
via the antifield cohomological perturbative method.
It is shown that the gauge algebra becomes abelian in the flat limit,
in contrast to what happens for totally symmetric gauge fields in AdS.
\end{abstract}
\newpage

\tableofcontents

\renewcommand{\thefootnote}{\arabic{footnote}}
\setcounter{footnote}{0}

\section{Introduction}\setcounter{equation}{0}

In the well-known papers \cite{Fradkin:1987ks,Fradkin:1986qy}, the
gravitational interaction problem (as well as self-interactions) for
higher-spin gauge fields was solved at the first nontrivial order by
going to a four-dimensional (anti-)de Sitter $(A)dS_4$ background.
This solved a longstanding problem and showed the importance of AdS
backgrounds. Subsequently, these results led to the solution of the
higher-spin interaction problem to all orders in interactions at the
level of field equations in the seminal papers
\cite{Vasiliev:1990en,Vasiliev:1992av,Vasiliev:2003ev}. The results
\cite{Fradkin:1987ks,Fradkin:1986qy} and
\cite{Vasiliev:1990en,Vasiliev:1992av,Vasiliev:2003ev} concern
higher-spin gauge fields which, when described in the metric-like or
Fronsdal formalism
\cite{Fronsdal:1978rb,Fang:1978wz,Fronsdal:1978vb,Fang:1979hq}, are given by
totally symmetric rank-$s$ tensors\footnote{For recent works on
cubic couplings among totally symmetric fields see
\cite{Vasiliev:2001wa, Alkalaev:2002rq, Metsaev:2005ar, Buchbinder:2006eq, Metsaev:2007rn,
Boulanger:2008tg, Sagnotti:2010at, Manvelyan:2010jr,
Polyakov:2010qs, Vasiliev:2011xx}.}. For a review of the key
mechanisms of higher-spin extensions of gravity, see
\cite{Bekaert:2010hw} while various reviews on Vasiliev's equations
can be found in
\cite{Vasiliev:2000rn,Vasiliev:2004qz,Bekaert:2005vh}.

Mixed-symmetry gauge fields are neither totally symmetric nor
totally antisymmetric ($p$-forms) and have first been described at
the Lagrangian level around flat background in
\cite{Curtright:1980yk,Aulakh:1986cb,Siegel:1986zi,Labastida:1987kw}.
For more recent works on mixed-symmetry fields in constantly curved
background see
\cite{deMedeiros:2003dc, deMedeiros:2003px, Alkalaev:2003qv,
Sagnotti:2003qa, Alkalaev:2005kw, Alkalaev:2006rw,
Bekaert:2006ix,Fotopoulos:2007nm,
Buchbinder:2007ix,Reshetnyak:2008gp,Skvortsov:2008sh,Zinoviev:2008ve,Campoleoni:2008jq,
Boulanger:2008up,Boulanger:2008kw, Alkalaev:2008gi,
Campoleoni:2009gs,Zinoviev:2009gh,Skvortsov:2009nv,Alkalaev:2009vm,
Skvortsov:2010nh}
and references therein.

As far as the problem of finding consistent interactions for
mixed-symmetry gauge fields is concerned, some analysis have been
done in flat background
\cite{Fradkin:1991iy,Bekaert:2002uh,Boulanger:2004rx,Bekaert:2004dz,Metsaev:2005ar,Metsaev:2007rn},
but in $\dSAdS$ background almost nothing has been achieved apart
from the very recent work \cite{Alkalaev:2010af} (see also the
earlier works \cite{Sezgin:2001zs,Sezgin:2001yf}). In
\cite{Zinoviev:2010av}, the electromagnetic interactions of massive
fields of the symmetry type studied in this paper have been studied
in the St\"uckelberg approach.

Generic irreducible mixed-symmetry gauge fields in $\AdS$ are very different
from their Minkowskian counterparts \cite{Brink:2000ag} in that they
possess only one differential gauge symmetry associated with a
single irreducible gauge parameter; not several at the same time,
like in flat spacetime. Gauge fields in $\AdS$ can be described within the
Alkalaev--Shaynkman--Vasiliev (ASV) approach \cite{Alkalaev:2003qv}
which is specific to $\AdS$ and presents the advantage of being
manifestly $\AdS$-covariant with a minimal set of off-shell fields.
However, the flat limit of the ASV formulation is not smooth in the sense of
non-conservation of dynamical degrees of freedom.

There is always, however, the possibility to reinstall all the
differential gauge symmetries for a generic mixed-symmetry gauge
fields in $\AdS$ at the price of adding extra fields, one for each
supplementary differential gauge parameter, which can be shifted to
zero at will provided the cosmological constant is nonvanishing
\cite{Brink:2000ag}, \cite{Boulanger:2008up, Boulanger:2008kw, Zinoviev:2009gh}. This is what we refer to as the
St\"uckelberg approach. Upon eliminating the extra fields of a
St\"uckelberg formulation, one arrives in $\AdS$ at the ASV
formulation.

It is the goal of the present paper to study in details the cubic
gravitational interaction problem in both flat and $AdS_d$
backgrounds for the simplest mixed-symmetry gauge field, \emph{i.e.}
one that is described in the metric-like formalism by a potential
whose Young symmetry is $\parbox{15pt}{\smallpic{\YoungBA}}$. Such a
field will here be called \emph{hook}, or $[2,1]$-type, gauge field.

We will treat the gravitational interactions of $[2,1]$-type gauge
field in $\AdS$ using a modified $\frac32$-order formalism and the
Fradkin--Vasiliev approach. These techniques will be applied in turn
to the St\"uckelberg and ASV formulations. In flat spacetime, we
will address the gravitational interaction problem in the
metric-like formalism and using the cohomological reformulation
\cite{Barnich:1993vg} of the consistent deformation procedure
\cite{Berends:1984rq}.

\vspace*{.2cm} The plan of the paper is as follows. After setting
the notation and conventions, in Section \ref{sec:CubicInteractions}
we review the methods of investigating cubic interactions. We recall
in Section \ref{sec:typepq} some results about the free $[p,q]$-type
gauge field in Minkowski and $\dSAdS$ backgrounds and give an
off-shell description of it in the frame-like St\"uckelberg and ASV
formalisms. Section \ref{sec:four} addresses the problem of cubic
interactions in $\AdS$. In Section \ref{sec:simplehookST} we
construct, from the St\"uckelberg vantage point, consistent cubic
gravitational interactions for the hook field in $\AdS$ (where
$d>4\,$) which contain the usual Lorentz-covariant minimal coupling
terms plus a finite sum of non-minimal terms, called
``quasi-minimal'' \cite{Boulanger:2008tg, Bekaert:2010hw} in the
context of totally symmetric gauge field in $\AdS\,$. The
St\"uckelberg action obtained therein allows a smooth flat limit.
The full expressions for the gauge transformations of the fields at
the first nontrivial order are explicitly given.
In Section \ref{sec:simplehookSTFV} the St\"uckelberg formulation is
treated using the Fradkin--Vasiliev construction. Then, in Section
\ref{sec:simplehookFV} we study the gravitational interactions within
the ASV formulation. We show that it agrees with the results of
Section \ref{sec:simplehookSTFV} upon partial gauge fixing of the
St\"uckelberg action. In Section \ref{sec:flatlimit}, using the
cohomological reformulation of the N\"other method
\cite{Barnich:1993vg,Henneaux:1997bm} and the results previously
obtained in \cite{Bekaert:2002uh,Boulanger:2004rx,Bekaert:2004dz},
we give the exhaustive list of cubic vertices in flat background
corresponding to the set of fields used in Sections
\ref{sec:simplehookST} and \ref{sec:simplehookSTFV}. We make contact
with the flat limit of the St\"uckelberg action. In particular, we
confirm that there are no possible nonabelian vertices in flat
space. In other words, switching on a cosmological constant enables
one to deform an abelian cubic action into a nonabelian one, in
sharp contrast to what happens for totally symmetric gauge fields
\cite{Boulanger:2008tg} where the nature of the gauge algebra is not
changed when going from $\AdS$ to flat background.
The conclusions are given Section \ref{sec:Conclusion}.
Finally, in Appendix \ref{sec:app} we review the metric-like St\"uckelberg formulation
for the free hook field in $\AdS{}$.

\section*{Notation and conventions}
Base manifold indices, or world indices, are denoted by Greek
letters $\mu,\nu,\ldots\,$, while Lorentz indices are denoted by
lower-case Latin letters. The Lorentz algebra $\mathfrak{so}(d-1,1)$
is associated with the metric $\eta_{ab}=\,$diag$(+,-,\ldots,-)$
where the indices $a,b,\ldots$ run over the values
$0,1,\ldots,d-1\,$.

A group of $p$ totally antisymmetric Lorentz indices will be denoted
by $[a_1a_2\ldots a_p]=a[p]\,$. Moreover, square brackets indicate
total antisymmetrization involving the minimal number of terms needed to
achieve antisymmetrization. To further simplify the notation we
will often use conventions whereby like letters imply complete
antisymmetrization, e.g. for $\phi^{u[2]}$
\be
\pl^{u}\phi^{uu}\equiv\pl^{[u}\phi^{uu]} =
\pl^{u_1}\phi^{u_2u_3}+\pl^{u_2}\phi^{u_3u_1}+\pl^{u_3}\phi^{u_1u_2}\quad.
\ee

The components of an irreducible $\mathfrak{gl}_d$ tensor whose
symmetry type consists of a Young diagram with two columns, the
first of length $p$ and the second of length $q\,$, will be denoted
$\varphi^{\mu[p], \nu[q]}\;$, $p\geqslant q\,$. The Young symmetry
described above is abbreviated by $\varphi \sim [p,q]\,$. The
components of a Lorentz tensor of type $[p,q]$ are denoted by
$\varphi^{a[p], b[q]}\,$. Note that, by abuse of notation, we do not
consider (anti) self-duality constraints on what we call Lorentz (or
$\AdS$) tensors in $d=2n\,$, so that a Lorentz tensor, in our
conventions, only obeys over-(anti)symmetrization and trace
constraints. The torsion-free, Lorentz-connection on the base
manifold is denoted by the symbol $\DL\,$. In flat background,
$(\DL)^2 \xi^a= 0\,$ for any vector $\xi^a$, whereas in $\AdS$ one
has $(\DL)^2 \xi^a=\Lambda h^a\wedge h_m \xi^m$, where $\Lambda$ is
a cosmological constant, $\Lambda<0$ for $\AdS\,$, and $h_{\mu}^{a}\,
dx^\mu\,$ are background vielbein one-forms. It is useful to set
$\lambda=\sqrt{|\Lambda|}$.

\section{Cubic interactions}
\setcounter{equation}{0} \label{sec:CubicInteractions}

\paragraph{Generalities.} Having a quadratic action $S_0[\Phi]$ that is invariant under
some abelian gauge transformations $\delta_0 \Phi$ as an input,
one may look for interaction vertices by expanding the nonlinear corrections
in powers of some formal coupling constant $g$
\begin{align}
{\cal S} & = {\cal S}_0 + g\,{\cal S}_1+... & \delta \Phi&=\delta_0 \Phi+g\,\delta_1 \Phi +{\cal O}(g^2)\quad.
\end{align}
The consistency condition $\delta {\cal S}=0$ at the leading nontrivial order,
which corresponds to cubic vertices, implies that
\begin{align}
\frac{\delta {\cal S}_1}{\delta \Phi} \;\delta_0 \Phi +
\frac{\delta {\cal S}_0}{\delta \Phi} \;\delta_1 \Phi =0 \quad.
\label{ConsistCond}
\end{align}
Noting that $\frac{\delta {\cal S}_0}{\delta \Phi}$ is the left-hand side of the
linear equations of motion, one can rewrite (\ref{ConsistCond}) as
\begin{align}
\label{onmassshellinv}
\left.\frac{\delta {\cal S}_1}{\delta \Phi}\; \delta_0
\Phi\rule{0pt}{12pt}\right|_{\frac{\delta {\cal S}_0}{\delta
\Phi}=0} =0
\end{align}
which is much easier to solve in practice. Given some solution ${\cal S}_1\,$,
the expression for the gauge transformations $\delta_1 \Phi$ can be extracted from (\ref{ConsistCond}).
In general $\delta_1 \Phi$ has a complicated form and is not needed
for most purposes.

\paragraph{Auxiliary fields, $\boldsymbol{1}$st-, $\boldsymbol{3/2}$-formalisms.}
In the first-order formalism which is widely used in higher-spin theory,
there are auxiliary fields which we denote collectively by $\Omega\,$.
The auxiliary fields $\Omega$ can be expressed, modulo gauge transformations,
in terms of physical fields $\Phi\,$.
Splitting (\ref{ConsistCond}) in terms of $\Phi$ and $\Omega$ gives
\begin{align}
\frac{\delta {\cal S}_1}{\delta \Phi} \;\delta_0 \Phi +
\frac{\delta {\cal S}_1}{\delta \Omega} \;\delta_0 \Omega +
\frac{\delta {\cal S}_0}{\delta \Phi} \;\delta_1 \Phi +
\frac{\delta {\cal S}_0}{\delta \Omega} \;\delta_1 \Omega = 0\quad.
\end{align}
On the one hand, one can use the $1$st-order formalism treating both
$\Phi$ and $\Omega$ as independent fields. But this requires a lot
of calculations including corrections to gauge transformations of
auxiliary field $\Omega\,$, which often turn out to be the most
complicated ones. Alternatively, in frame-like formulation of
gravity and supergravity there is the well-known $\frac{3}{2}$-order
formalism where one takes into account the variations of physical
fields $\Phi$ only, all the calculations being done on the solutions
of the complete algebraic equations for the auxiliary field $\Omega$
\begin{align}
\left[ \frac{\delta {\cal S}_1}{\delta \Phi} \;\delta_0 \Phi +
\frac{\delta {\cal S}_0}{\delta \Phi} \;\delta_1 \Phi
\right]_{\frac{\delta ({\cal S}_0 + {\cal S}_1)}{\delta \Omega} = 0}
= 0\quad.
\end{align}
The advantage is that there is no need to consider the corrections
$\delta_1\Omega$ to the $\Omega$ gauge transformations. However, one
has to solve non-linear equations for $\Omega$ and this can be
highly nontrivial. In Section \ref{sec:simplehookST} we use a
modified $\frac{3}{2}$-order formalism very well suited for the
investigations of cubic vertices
\begin{align}\label{firstorderform}
\left[ \frac{\delta {\cal S}_1}{\delta \Phi} \;\delta_0 \Phi +
\frac{\delta {\cal S}_1}{\delta \Omega} \;\delta_0 \Omega +
\frac{\delta {\cal S}_0}{\delta \Phi} \;\delta_1 \Phi
\right]_{\frac{\delta {\cal S}_0}{\delta \Omega} = 0} = 0\quad.
\end{align}
Here also there is no need to consider $\delta_1\Omega$ and we have to make all calculations on the
solutions of free $\Omega$ field equations only. If one is not interested in $\delta_1\Phi\,$,
then the equation (\ref{onmassshellinv}) can be used to find ${\cal S}_1\,$.

\paragraph{Fradkin--Vasiliev cubic interactions, \cite{Fradkin:1986qy}.}
It is very convenient for the purpose of finding interactions to
reformulate a field theory in the unfolded form
\cite{Vasiliev:1988xc},
\begin{align}
&R^\aA=0, && R^\aA=d W^\aA-F^\aA(W), && F^\aB\frac{\pl
F^\aA}{\pl W^\aB}\equiv0, \label{unfld}
\end{align}
so that (i) the fields $W^\aA$ form a set of differential forms of
some degrees; (ii) all exterior derivatives of fields are expressed
in terms of fields themselves, moreover $F^\aA(W)$ is a function
built with the use of exterior product of fields only; (iii)
$d^2\equiv0$ implies an integrability condition for $F^\aA(W)$. In
other words the set $W^\aA$ is closed under the de Rham derivative
and there are no other relations between $dW^\aA$ then those given
by $F^\aA$. Then (\ref{unfld}) ensures the gauge symmetry
\begin{align}
\delta W^\aA= d\xi^\aA+\xi^\aB\frac{\pl F^\aA}{\pl W^\aB}
\end{align}
where the first term is absent if the form degree of $W^\aA$ is zero.

The anti-de Sitter background itself can be thought of as a part of the unfolded system of equations:
\begin{align}
d h^a+\varpi\fud{a,}{c}\wedge h^c&=0\quad ,\label{zerocurvGA}\\
d\varpi^{a,b}+\varpi\fud{a,}{c}\wedge\varpi^{c,b}&=\Lambda h^a\wedge
h^b\quad .
\label{zerocurvGB}
\end{align}
In what follows we shall not use the full system of unfolded
equations to describe a dynamical field of some spin, but only
several Yang--Mills-like curvatures $R^\aA$ that are relevant for
the cubic action as they contribute to the quadratic action.
The strategy is
\begin{itemize}
 \item[(1)] for a
required multiplet of gauge fields, for which we would like to
investigate cubic interactions, one has to give unfolded curvatures
$R^\aA_0$ that are linear in gauge fields,
\begin{align}
R^\aA_0&=\DL W^\aA+F^\aA_\aB(h)W^\aB
\end{align}
but could be nonlinear in the background fields as manifested in
$F^\aA_\aB(h)\,$. Note that $\varpi$ can appear only as a part of the
Lorentz covariant derivative $\DL=d+\varpi\,$.
The curvature $R^\aA_0$ can be read off from
\cite{Lopatin:1987hz, Vasiliev:2001wa, Alkalaev:2003qv,
Skvortsov:2008vs, Zinoviev:2009gh, Skvortsov:2009zu,
Zinoviev:2009wh, Zinoviev:2008ve, Boulanger:2008up,Boulanger:2008kw,
Zinoviev:2008ze}. The indices $\aA,\aB,...$ run over certain set of
irreducible Lorentz tensors. As the number of gauge forms $W^\aA$
for some particular field is finite $R^{\aA}_0=0$ cannot describe
propagating fields. The way out is that not all of the curvatures
$\DL W^\aA+F^\aA_\aB(h)W^\aB$ are zero, some being proportional to
zero-forms $C^\ba$, called generalized Weyl tensors:
\begin{align}
R^\aA_0&=F^{\aA}_{\ba}(h)C^\ba\label{unfldwithweyls}\,,
\end{align}
which are consistent unfolded equations provided that $\DL C^\ba$
satisfy their own equations.
In what follows the curvatures for $C^\ba$ are not needed;
\item[(2)] find a quadratic action of the form
\begin{align}
S=\frac12\sum_{\aA,\aB}\int I_{\aA\aB; u...u}\, R^\aA_0\wedge
R^\aB_0\wedge h^u\wedge...\wedge h^u,
\end{align}
which is gauge invariant by construction. The $I_{\aA\aB; u...u}$ are
some invariant tensors built out of $\epsilon_{ab...c}$ and $\eta_{ab}$.
The action contains some free coefficients, which are generally
responsible for normalization of actions for individual constituents
of the multiplet and for a freedom to add boundary terms;
\item[(3)]  to extend the curvatures $R^\aA_0$
\begin{align}
R^\aA&=\DL W^\aA+F^\aA_\aB(h)W^\aB+\frac12 gF^{\aA}_{\aB\aC}(h)W^\aB\wedge W^\aC
\end{align}
with terms quadratic in the fields while maintaining the integrability condition
(\ref{unfld}) to the order $g$, which implies
\begin{align}\label{deltaR}
\delta R^\aA&=\xi^\aC R^\aD_0 \frac{\pl F^\aA}{\pl W^\aC
\pl W^\aD}+O(g^2)=g\xi^\aC R^\aD_0 F^{\aA}_{\aC\aD}+O(g^2)\,,
\end{align}
where we have replaced $R^\aD$ with $R^\aD_0$ on right-hand side as
$F^{\aA}_{\aC\aD}$ has already brought in one power of $g$. The coefficients
$F^{\aA}_{\aC\aD}$ are in fact the structure constants of some
higher-spin algebra \cite{Fradkin:1986ka, Vasiliev:2004cm}, so that
$R=dW+W\star W\,$. In the present paper we do not consider the full higher-spin
algebra as we look for some particular cubic vertices and
do not know $F^{\aA}_{\aC\aD}$ for all generators;
\item[(4)] to insert the
corrected curvatures $R^\aA$ into the action instead of the
linearized $R^{\aA}_0$ and to adjust free coefficients such that the
action is gauge invariant to the order $g$:
\begin{align}\label{gaugevariationoff}
\delta S=&\sum_{\aA,\aB}\int  I_{\aA\aB; u...u}\, \delta R^\aA\wedge
R^\aB\wedge h^u\wedge...\wedge h^u=\\=&g\sum_{\aA,\aB}\int
I_{\aA\aB; u...u}\,  \xi^\aC\wedge R^\aD_0 F^{\aA}_{\aC\aD} \wedge
R^\aB_0\wedge h^u\wedge...\wedge h^u+O(g^2)\quad.
\end{align}
\end{itemize}
At this stage one may use a modified $\frac32$-order formalism
(\ref{firstorderform}) or, if there is no need in finding
$\delta_1\Phi$, one can apply a modified (\ref{onmassshellinv}),
\begin{align}\label{onmassshellinvm}
\left[\frac{\delta {\cal S}_0}{\delta \Phi} \delta^g_1
\Phi+\frac{\delta {\cal S}_1}{\delta \Phi} \delta_0
\Phi\rule{0pt}{12pt}\right]_{\frac{\delta {\cal S}_0}{\delta
\Phi}=0} =0
\end{align}
where $\delta_1=\delta^{ex}_1+\delta^g_1$ and $F^{\aA}_{\aB\aC}$ are
responsible for $\delta^g_1$, which appears naturally and makes the
expression in brackets more simple (\ref{deltaR}). In this case
there are additional simplifications due to the fact that most of
$R^\aA_0$ are zero on-shell, a few being equal to Weyl tensors
(\ref{unfldwithweyls}). Applying (\ref{unfldwithweyls}) one reduces
(\ref{gaugevariationoff}) to
\begin{align}
\delta S=g\sum_{\aA,\aB}\int I_{\aA\aB; u...u}  C^\ba C^\bb
F^{\aA}_{\aC\aD}\,\, \xi^\aC\wedge F^{\aD}_{\ba}(h)  \wedge
F^{\aB}_{\bb}(h)\wedge h^u\wedge...\wedge h^u+O(g^2)=0,
\end{align}
which is a purely algebraic problem of adjusting free coefficients
in order for various combinations of Weyl tensors to cancel each
other.

Actually, in the original paper \cite{Fradkin:1986qy}\footnote{For
some reviews, see e.g
\cite{Vasiliev:2000rn,Vasiliev:2004cp,Vasiliev:2004qz}.} the
coefficients $F^{\aA}_{\aC\aD}$ were completely known for the
multiplet of totally symmetric fields of spins $s=0,1,...$, in
contrast to the present paper where we just probe some of
$F^{\aA}_{\aC\aD}$ for the $\mathfrak{osp}(1|2)$ higher-spin algebra
of \cite{Vasiliev:2004cm} describing certain mixed-symmetry fields
in addition to totally-symmetric ones.

\section{Type$\,$-$\boldsymbol{[p,q]}$ gauge fields}
\setcounter{equation}{0} \label{sec:typepq}

In this section we present a frame-like formulation for type-$[p,q]$
fields. The cases $[1,1]$ and $[2,1]$ will be treated in details in
the rest of the paper.

\paragraph{Minkowski space.}
According to \cite{Skvortsov:2008sh, Skvortsov:2008vs}, the unfolded
formulation for type-$[p,q]$ field in Minkowski space starts with
two fields
\begin{align}
\mbox{type }\Ya{p,q}:& &&e^{a[q]}\fm{p}&&\omega^{a[p+1]}\fm{q}
\label{secondset}
\end{align}
as a vielbein and spin-connection spin-$\Ya{p,q}$ field, where the
subscripts on the above fields indicate their respective
differential form degrees. One can construct linearized unfolded
curvatures for these fields as
\begin{align}
R^{a[q]}\fm{p+1}&=d e^{a[q]}\fm{p}-h_{m[p-q+1]}\omega^{a[q]m[p-q+1]}\fm{q}\,,\label{unfldcurA}\\
R^{a[p+1]}\fm{q+1}&=d \omega^{a[p+1]}\fm{q}\label{unfldcurB}\,.
\end{align}
On-mass-shell the curvatures obey
\begin{align}
R^{a[q]}\fm{p+1}&=0\,,\\
R^{a[p+1]}\fm{q+1}&=h_{m[q+1]} C^{a[p+1],m[q+1]}\fm{0}\,,
\end{align}
where $C^{a[p+1],m[q+1]}\fm{0}$ is the Weyl tensor. The quadratic
action has the form
\begin{align}
S_{p,q}&=\frac12 \int \left(R^{u[q]}\fm{p+1} \wedge
\omega^{u[p+1]}\fm{q} +(-1)^{p+1} \, e^{u[q]}\fm{k}\wedge
R^{u[p+1]}\fm{q+1}\right)H_{u[p+q+1]} \,, \label{ActionKQB}
\end{align}
where we have used volume forms $H_{u[k]}$
\begin{align*}
H_{u[k]}&=\epsilon_{u_1...u_k b_{k+1}...b_{d}}\, h^{b_{k+1}}\wedge
...\wedge h^{b_{d}}
\end{align*}
which form a basis set of $(d-k)$ forms, resulting in the identity
\begin{align}
\label{MSIdentityVielbeins} h^c\wedge H_{u_1...u_k } \;=\;
\frac{(-1)^k}{(d-k+1)}\;\sum_{i=1}^{i=k}(-1)^{i}\delta^c_{u_i}\,H_{u_1...\hat{u}_i...u_k}
\quad\,.
\end{align}

However, it turns out that this set of fields/curvatures does not
admit a straightforward deformation to the anti-de Sitter space.

\paragraph{St\"uckelberg formulation.}
The BMV conjecture \cite{Brink:2000ag},
proved in \cite{Boulanger:2008up, Boulanger:2008kw, Alkalaev:2009vm},
states that an irreducible \AdS{} massless field decomposes into a direct sum of
irreducible Minkowski massless fields in the flat limit $\Lambda\rightarrow0\,$.
For a unitary spin-$[p,q]$ field the decomposition is
\begin{equation}
\left.\mbox{AdS:\quad unitary spin-}
\Ya{p,q}\phantom{\rule{0pt}{16pt}}\right|_{\Lambda\rightarrow0}
\;\cong\;\; \mbox{spin-}\Ya{p,q}\oplus \mbox{spin-}\Ya{p-1,q}\quad
\mbox{: Minkowski}. \label{TwocolumnsFlatLimit}
\end{equation}
This can be understood as follows. A spin-$[p,q]$ massless field in
Minkowski space possesses two independent differential gauge symmetries
with parameters having the types $[p-1,q]$ and $[p,q-1]\,$.
The first gauge symmetry gets broken in \AdS{}; its role was to
remove, from the components of the spin-$[p,q]$ field,
the spin-$[p-1,q]\,$ polarization.
The spin-$[p,q-1]$ gauge parameter is still activated in \AdS{}.
Thus, if one wants the total number of physical degrees of freedom to be a
conserved quantity in the limit $\Lambda\rightarrow0\,$,
then the spin-$[p-1,q]$ polarization has to decouple and will become
an independent massless field.
The above scenario is what one expects from representation theory:
in the flat limit an irreducible representation of \ads{} becomes an
$\iso$-reducible one.

That \iso{} is not semisimple indeed results in a drastic difference between
formulations in \AdS{} and Minkowski backgrounds.
However, from a given field-theoretical formulation in a flat space in one higher dimension,
one can obtain all the information one wishes about the corresponding massive or
massless field in \AdS{} or flat backgrounds.
This is a powerful technique, systematically applied in various cases by one
of the authors
\cite{Zinoviev:2001dt,Zinoviev:2003ix,Zinoviev:2008ve,Zinoviev:2008ze,Zinoviev:2009gh,Zinoviev:2009vy}
and that can be used in order to derive the Lagrangian formulation of
any given field (massive or massless, in \AdS{} or Minkowski):
Starting from the maximal St\"uckelberg description of a generically massive field
in \AdS{} background, one has at one's disposal all the fields needed to describe its
massless $m^2\rightarrow0$ and/or flat $\Lambda\rightarrow0$ limit without losing any
degrees of freedom.
For example, given a totally symmetric spin-$s$ massive field,
its massless limit produces a set of massless fields
whose spectrum of spins is obtained by a dimensional reduction of length-$s$ one-row
Young diagram $[1,1,\ldots,1]$ of \msv{} to various one-row diagrams of \mls with less
and less cells.
Hence, in the simplest case of a spin-$s$ massive field one may take a set of fields
that is used to describe massless fields of spins $0,1,...,s$.
The massive Lagrangian is then a sum of Lagrangians of massless fields supplemented by
various mixing terms with $1$ and $0$ derivatives. The procedure is quite natural and has
already been tested both in Minkowski and \AdS{} for all fields whose spin is given by an
arbitrary two-row Young diagram \cite{Zinoviev:2009gh}.

This  idea was also mentioned in \cite{Biswas:2002nk} and has been used
\cite{Boulanger:2008up,Boulanger:2008kw} in order to describe in a
geometric way massive and massless \AdS{} fields of arbitrary symmetry type,
starting from the dimensional reduction of the Minkowskian $(d+1)$-dimensional
geometric formulation \cite{Skvortsov:2008sh, Skvortsov:2008vs} developed by one of the
authors.
In that way, it was possible to reproduce and understand the Brink--Metsaev--Vasiliev
(BMV) pattern of St\"uckelberg fields and their possible decouplings,
at any  given value of the mass parameter.
For a very recent and interesting works related to \cite{Biswas:2002nk},
see e.g. \cite{Alkalaev:2011zv, Grigoriev:2011gp} and references therein.

The advantage (and definition) of the maximal St\"uckelberg formulation is that all limits
(massless and/or flat) are smooth and all the fields have a gauge transformation.
For a massive field in $\AdS$, at critical values of $m^2$ certain mixing coefficients in
the Lagrangian go to zero and the latter splits into two pieces,
one describing a massless (or partially massless) field in $\AdS$ which still has more
degrees of freedom than the Minkowski massless field with the same spin.
If, instead of choosing a critical value for $m^2$ in the maximal St\"uckelberg
Lagrangian, one takes the limit $\Lambda\rightarrow0\,$, then all the mixing coefficients
go to zero and one obtains a direct sum of massless Lagrangians, one for each of the
fields appearing in the initial Lagrangian.

\paragraph{St\"uckelberg formulation for type-$\boldsymbol{[p,q]}$ fields.}
According to (\ref{TwocolumnsFlatLimit}) in order to construct St\"uckelberg formulation for type-$\boldsymbol{[p,q]}$ field one may take
\begin{align}
\mbox{type }\Ya{p-1,q}:& &&e^{a[q]}\fm{p-1}&&\omega^{a[p]}\fm{q}\,,\\
\mbox{type }\Ya{p,q}:& &&e^{a[q]}\fm{p}&&\omega^{a[p+1]}\fm{q}\,.
\label{secondsetbis}
\end{align}
The ansatz for curvatures contains all mixing terms reads
\begin{align}
R^{a[q]}\fm{p}&=\DL e^{a[q]}\fm{p-1}-h_{m[p-q]}\omega^{a[q]m[p-q]}\fm{q}-\alpha e^{a[q]}\fm{p}\,,\label{curvA}\\
R^{a[p]}\fm{q+1}&=\DL \omega^{a[p]}\fm{q}-\beta h_{m}\omega^{a[p]m}\fm{q}\,,\label{curvB}\\
R^{a[q]}\fm{p+1}&=\DL e^{a[q]}\fm{p}-h_{m[p-q+1]}\omega^{a[q]m[p-q+1]}\fm{q}-\gamma h^ah_m e^{a[q-1]m}\fm{p-1}\,,\label{curvC}\\
R^{a[p+1]}\fm{q+1}&=\DL \omega^{a[p+1]}\fm{q}-\delta
h^a\omega^{a[p]}\fm{q}\,.\label{curvD}
\end{align}
The integrability condition (\ref{unfld}) implies that
\begin{align}
&\alpha=\frac{\Lambda}{\delta} (-)^q\,, && \gamma=-\delta\,, &&
\beta=\frac{\Lambda}{\delta} (-)^{p-1}\label{coefssolution}\,.
\end{align}
Choosing $\delta \sim \sqrt{|\Lambda|}\equiv \lambda$, one can have
a smooth flat limit $\Lambda\rightarrow0$, with the four curvatures
decoupling into two independent sets of
(\ref{unfldcurA})-(\ref{unfldcurB}). On-mass-shell we have
\begin{align}
R^{a[q]}\fm{p}&=0\,,&
R^{a[p]}\fm{q+1}&=h_{m[q+1]}C^{a[p],m[q+1]}\,,\label{onmassshellpqA}\\
R^{a[q]}\fm{p+1}&=0\,,&
R^{a[p+1]}\fm{q+1}&=h_{m[q+1]}C^{a[p+1],m[q+1]}\,,\label{onmassshellpqB}
\end{align} featuring two Weyl tensors in accordance with the BMV
conjecture, \cite{Brink:2000ag}.

The action, which is valid both in Minkowski and $AdS$ is simply a
sum of (\ref{ActionKQB}), where the curvatures are to be replaced
with (\ref{curvA})-(\ref{curvD}).
\begin{align}
S&=S_{p-1,q}+\frac{-\Lambda}{(p+1)\delta^2}S_{p,q}\,,
\end{align}
where the relative coefficient is fixed by gauge invariance. Note
that the action does not have a manifestly gauge invariant form.
This can be cured in $AdS$.

In anti-de Sitter space one can indeed cast the action into a
manifestly gauge invariant form
\begin{align}
S=a_1&\int{R\fm{q+1}}^{u[q+1]a[p-q-1]}\wedge {R\fm{q+1}}\fud{u[q+1]}{a[p-q-1]}\wedge H_{u[2q+2]}+ \nonumber \\
+a_2&\int{R\fm{q+1}}^{u[q+1]a[p-q]}\wedge {R\fm{q+1}}\fud{u[q+1]}{a[p-q]}\wedge H_{u[2q+2]}+ \nonumber\\
+a_3&\int{R\fm{p}}^{u[q]}\wedge {R\fm{q+1}}^{u[p+1]}\wedge H_{u[p+q+1]}+ \nonumber \\
+a_4&\int\DL\left({R\fm{q+1}}^{u[q+1]a[p-q-1]}\wedge
{R\fm{q+1}}\fud{u[q+2]}{a[p-q-1]}\wedge H_{u[2q+3]}\right)\,,
\end{align}
where the last term is a boundary term and allows to set $a_1/a_2$
at will, then we may put $a_4=0$. The action is manifestly gauge invariant. However, if one
expands the first three terms one finds contributions of the form
${\omega\fm{q}}^{u[q]a{p-q}}\wedge\DL{\omega\fm{q}}\fud{v[p+1]}{a[p-q]}H_{u[q]v[p+1]}$
that are of the third order in derivatives upon solving equations
for $\omega$'s in terms of tetrad-like fields $e$. The requirement
for higher-derivative terms to vanish gives one constraint on
$a_1,a_2,a_3$
\begin{align}
-\frac{2(q+1)}{d-2q-1}\left(a_1\frac{\Lambda}{\delta}+
a_2(d-p-q-1)\delta\right)+a_3(-)^{p+q+1}\frac{(p+1)!(d-p-q-1)!}{(q+1)!(d-2q-1)!}=0\,.
\end{align}

\paragraph{Interlude.} One \cite{Stelle:1979aj, Vasiliev:2001wa} can collect
$\varpi^{a,b}$ and $h^a$ as different components of a single
\ads-connection $\Omega^{A,B}\equiv -\Omega^{B,A} \equiv
\Omega^{A,B}_\mu dx^\mu$, $A,B,...=0...d$, where the decomposition
of $\Omega^{A,B}$ into a $d\times d$ antisymmetric matrix
$\varpi^{a,b}$ and a Lorentz vector $h^a$ can be performed in a
\ads-covariant way by introducing a normalized vector field $V^A$,
$V^C V_C=1$, called compensator, \cite{Vasiliev:2001wa},
\begin{align}
H^A & = \lambda \,\DO V^A \quad, & \Omega_L^{A,B} = \;\Omega^{A,B}+
\lambda(V^AH^B-H^AV^B)\quad , \label{AdSDefinitions}
\end{align}
where $\DO=d+\Omega$, $(\DO)^2=0$, which is equivalent to
(\ref{zerocurvGA})-(\ref{zerocurvGB}). The \ads-covariant
definitions for $h^a$ and $\varpi^{a,b}$ are given by $H^A$ and
$\Omega_L^{A,B}$, with the Lorentz covariant derivative being
$\DL=d+\Omega_L$, which is manifested in
\begin{align}
H^BV_B & = 0 \quad,   &   DV^A  & = 0 \quad ,  &   DH^A & = 0 \quad
\end{align}
and the fact that, choosing $V^A=\delta^A_{d}$ ($d$ means the value
of the index rather than an index), one recovers
\begin{align}
H^a_\mu & = \lambda\,{\Omega_\mu}^{a,}_{\phantom{a,}d} \quad, &
H^d_\mu & = 0 \quad, & \Omega_L^{a,b} & = \Omega^{a,b}\quad.
\end{align}

\paragraph{Manifestly $\boldsymbol{AdS}$-covariant formulation.}
One can go further and construct a formulation for type-$[p,q]$
fields that in addition to being manifestly gauge invariant is also
manifestly covariant under global symmetries of $AdS$,
\cite{Vasiliev:2001wa, Alkalaev:2003qv, Alkalaev:2003hc,
Alkalaev:2005kw, Alkalaev:2006rw}.

One first notes that with the help of gauge parameter
$\xi^{a[q]}\fm{p-1}$ of $e^{a[q]}\fm{p}$
\begin{align}
\delta e^{a[q]}\fm{p-1}= \DL
\xi^{a[q]}\fm{p-2}+(-)^{p-q}h_{m[p-q]}\xi^{a[q]m[p-q]}\fm{q}+\alpha
\xi^{a[q]}\fm{p-1}\label{gaugetrA}
\end{align}
one can gauge away $e^{a[q]}\fm{p-1}$ completely. It is obvious that
$\omega^{a[p]}\fm{q}$ and $e^{a[q]}\fm{p}$ have the same number of
components. In the gauge where $e^{a[q]}\fm{p-1}=0$, the
zero-curvature condition on (\ref{curvA}) implies that
\begin{align}
h_{m[p-q]}\omega^{a[q]m[p-q]}\fm{q}-\alpha e^{a[q]}\fm{p}=0\,,
\end{align}
i.e. $e^{a[q]}\fm{p-1}$ is just an avatar for $\omega^{a[p]}\fm{q}$.
The curvature (\ref{curvC}) then does not carry any new information
and can be abandoned to the benefit of (\ref{curvB}). The resulting
formulation is based on two fields $\omega^{a[p]}\fm{q}$ and
$\omega^{a[p+1]}\fm{q}$ with the unfolded curvatures of the form
\begin{align}
R^{a[p]}\fm{q+1}&=\DL \omega^{a[p]}\fm{q}+\lambda h_{m}\omega^{ma[p]}\fm{q}\label{curvAA}\,,\\
R^{a[p+1]}\fm{q+1}&=\DL \omega^{a[p+1]}\fm{q}-\lambda
h^a\omega^{a[p]}\fm{q}\,,\label{curvAB}
\end{align}
where we have made the choice $\delta=\sqrt{|\Lambda|}\equiv
\lambda$. The action now reduces to three terms
\begin{align}
S=a_1&\int{R\fm{q+1}}^{u[q+1]a[p-q-1]}\wedge {R\fm{q+1}}\fud{u[q+1]}{a[p-q-1]}\wedge H_{u[2q+2]}+\nonumber\\
+a_2&\int{R\fm{q+1}}^{u[q+1]a[p-q]}\wedge {R\fm{q+1}}\fud{u[q+1]}{a[p-q]}\wedge H_{u[2q+2]}+\nonumber\\
+a_4&\int\DL\left({R\fm{q+1}}^{u[q+1]a[p-q-1]}\wedge
{R\fm{q+1}}\fud{u[q+2]}{a[p-q-1]}\wedge H_{u[2q+3]}\right)
\end{align}
with no restriction on $a_1,a_2,a_4$. The boundary term again serves \cite{Alkalaev:2003hc, Alkalaev:2003qv}
as a tool to adjust $a_1/a_2$ at will. The on-mass-shell condition
(\ref{onmassshellpqA})-(\ref{onmassshellpqB}) reduces to
\begin{align}
R^{a[p]}\fm{q+1}&=h_{m[q+1]}C^{a[p],m[q+1]}\,,\\
R^{a[p+1]}\fm{q+1}&=h_{m[q+1]}C^{a[p+1],m[q+1]}\,.\label{curvBB}
\end{align}

Now one can realize (\ref{curvAA})-(\ref{curvAB}) as two projections
$R^{a[p+1]}\fm{q+1}$ and $R^{a[p+1]M}\fm{q+1}V_M$ of a single
curvature $R^{A[p+1]}\fm{q+1}=\DO W^{A[p+1]}\fm{q}$ for a
generalized \ads-connection $W^{A[p+1]}\fm{q}$, which has an
analogous decomposition into two generalized Lorentz connections to
be identified with $\omega^{a[p]}\fm{q}$ and
$\omega^{a[p+1]}\fm{q}$. The action can also be rewritten in a
\ads-covariant form \cite{Alkalaev:2003hc, Alkalaev:2003qv},
\begin{align}
S=a_1&\int{R\fm{q+1}}^{U[q+1]A[p-q-1]M}V_M\wedge {R\fm{q+1}}\fudu{U[q+1]}{A[p-q-1]}{N}V_N\wedge H_{U[2q+2]}+\nonumber\\
+a_2&\int{R\fm{q+1}}^{U[q+1]A[p-q]}\wedge {R\fm{q+1}}\fud{U[q+1]}{A[p-q]}\wedge H_{U[2q+2]}+\nonumber\\
+a_4&\int\DO\left({R\fm{q+1}}^{U[q+1]A[p-q-1]M}V_M\wedge
{R\fm{q+1}}\fud{U[q+2]}{A[p-q-1]}\wedge H_{U[2q+3]}\right)
\end{align}
with
\begin{align*}
H_{U[k]}&=\epsilon_{U_1...U_k B_{k+1}...B_{d}W}\, H^{B_{k+1}}\wedge
...\wedge H^{B_{d}} V^W\,. &
\end{align*}
Such formulation in terms of generalized connections of the anti-de
Sitter algebra is referred to as ASV formulation due to
\cite{Alkalaev:2003qv}, where it was introduced first, see
\cite{Alkalaev:2005kw,Alkalaev:2006rw,Boulanger:2008up,Boulanger:2008kw,Skvortsov:2009zu,Skvortsov:2009nv}
for developments and generalizations. Within the ASV formulation a
set of frame-like fields is organized in a compact way as various
projections of a single generalized \ads-connection. However, this
is achieved at the price of losing St\"uckelberg symmetries and
associated fields that make the flat limit smooth. Therefore, ASV
formulation is restricted to $AdS$ and has a singular flat limit,
with the Van Dam--Veltman--Zakharov-like discontinuity in the number
of physical degrees of freedom.

\section{Gravitational interactions of type$\,$-$[2,1]$ fields in
$\boldsymbol{AdS}$}\label{sec:four} \setcounter{equation}{0} In this
section we are going to present three different ways of constructing
gravitational interactions for the $[2,1]$-type gauge fields in
\AdS.

\subsection{St\"uckelberg formulation and $\boldsymbol{\frac32}$-approach}
\label{sec:simplehookST} Firstly we apply a modified
$\frac32$-formalism in the presence of maximal set of St\"uckelberg
fields.

\paragraph{Kinematics.} In accordance with section
\ref{sec:typepq} we will use the following fields for description of
hook: two form $\Phi^a$ and one forms $\Omega^{a[3]}$,
$\Omega^{a[2]}$ and $f^a$, leaving notations $e^a$ and
$\omega^{a[2]}$ for the description of graviton. In this notation
the free Lagrangian for a hook in $AdS$ can be written as follows
\begin{eqnarray}
{\cal L}_0 &=& [ - \frac{1}{2} \Omega^{a[3]} h_m
h_m \Omega^{am[2]} + \Omega^{a[3]} D \Phi^a ] H_{a[4]} + \nonumber \\
 && + [ \frac{1}{2} \Omega^{a[2]} h_m \Omega^{am} -
\Omega^{a[2]} D f^a ] H_{a[3]} + \nonumber \\
 && + m [ h_m \Omega^{a[2]m} f^a + \Omega^{a[2]} \Phi^a ]
 H_{a[3]}\,,
\end{eqnarray}
where $m^2 = 3\lambda^2$. It is invariant under the following set of
gauge transformations
\begin{eqnarray}
\delta_0 \Phi^a &=& D z^a + h_m h_m \chi^{m[2]a} + \frac{m}{3} h^a
h_m
\xi^m\,,  \\
\delta_0 \Omega^{a[3]} &=& D \chi^{a[3]} + \frac{m}{3} h^a
\chi^{a[2]}\,, \\
\delta_0 f^a &=& D \xi^a + h_m \chi^{ma} + m z^a\,, \\
\delta_0 \Omega^{a[2]} &=& D \chi^{a[2]} - m h_m \chi^{ma[2]}\,.
\end{eqnarray}
Correspondingly, we can construct four gauge invariant objects
(Yang--Mills-like curvatures)\footnote{To make connection with
(\ref{coefssolution}), $\alpha=m$, $\beta=m$, $\gamma=m/3$,
$\delta=-m/3$, $p=2$, $q=1$.}
\begin{eqnarray}
R^a &=& D \Phi^a - h_m h_m \Omega^{m[2]a} - \frac{m}{3} h^a h_m f^m\,, \\
R^{a[3]} &=& D \Omega^{a[3]} + \frac{m}{3} h^a \Omega^{a[2]}\,, \\
K^a &=& D f^a + h_m \Omega^{ma} - m \Phi^a\,, \\
K^{a[2]} &=& D \Omega^{a[2]} - m h_m \Omega^{ma[2]}\,.
\end{eqnarray}
They satisfy the following differential identities
\begin{eqnarray}
D R^{a[3]} &=& - \frac{m}{3} h^a K^{a[2]}, \qquad
D R^a = - h_m h_m R^{m[2]a} - \frac{m}{3} h^a h_m K^m\,, \\
D K^{a[2]} &=& m h_m R^{m[a[2]}, \qquad D K^a = - h_m K^{ma} - m
R^a\,.
\end{eqnarray}
Note here that on the solutions of the equations for the auxiliary
fields $\Omega^{a[3]}$ and $\Omega^{a[2]}$ we have
\begin{equation}
R^a = 0, \quad K^a = 0 \quad \Longrightarrow \quad h_m h_m R^{m[2]a}
= 0, \quad h_m K^{ma} = 0\,.
\end{equation}

\paragraph{Minimal interactions.}\label{subsec:minimal}

To illustrate how our modified formalism work let us begin with the
free Lagrangian for a massless hook in a flat spacetime and
corresponding initial gauge transformations, where now $D^2=0$,
\begin{equation}
{\cal L}_0 = \frac{1}{2} [ - h_m h_m \Omega^{m[2]a} \Omega^{a[3]} +
2 \Omega^{a[3]} D \Phi^a ] H_{a[4]}\,,
\end{equation}
\begin{equation}
\delta_0 \Omega^{a[3]} = D \chi^{a[3]}, \qquad \delta_0 \Phi^a = D
z^a + h_m h_m \chi^{m[2]a}\,.
\end{equation}
The most general ansatz for a cubic vertex with two derivatives has
the form (here we put the gravitational coupling constant to 1)
\begin{equation}
{\cal L}_1 = [ a_1 e_m h_m \Omega^{m[2]a} \Omega^{a[3]} + a_2 e^a
h_m \Omega^{m[2]a} \Omega_m{}^{a[2]} - \Omega^{a[3]} \omega^a{}_m
\Phi^m ] H_{a[4]}\,.\label{twoderivvertex}
\end{equation}
Let us consider gauge transformations for the graviton, first,
\begin{align}&\delta e^a = D \eta^a + h_m \eta^{ma}, && \delta \omega^{a[2]} = D
\eta^{a[2]}\,.\end{align}  Direct calculations show that the
variation of ${\cal L}_1$ under the above transformations cancels
provided that $a_1 = \frac{3}{2}$ and $a_2 = - \frac{3}{2}$,
together with appropriate corrections to gauge transformations
\begin{equation}
\delta_1 \Phi^a = \eta^{am} \Phi_m + \eta_m h_m \Omega^{m[2]a}\,.
\end{equation}
So we have fixed all the coefficients in the cubic vertex and we
still have to consider the gauge variation ${\cal L}_1$ under
$\delta_0 \Phi^a$. It is easy to check that it is impossible to
achieve complete invariance under these transformations. The best
possible result is obtained with the following corrections to
$\delta_1 \Phi^a$
\begin{equation}
\delta_1 \Phi^a = - e_m h_m \chi^{m[2]a} - \omega^{am} z_m\,.
\end{equation}
This leaves us with the residual terms
\begin{equation}
res_{01} = [ \chi^{a[3]} R^{am} \Phi_m - \Omega^{a[3]} R^{am} z_m ]
H_{a[4]} \label{res01}\,,
\end{equation}
that are directly related to the fact that covariant derivatives do
not commute.

Similarly, we repeat the above procedure in the sector of the
St\"uckelberg spin-$2$ field, giving
\begin{align}
\delta_1 f^a &= -\omega\fud{a}{m} \xi^m +\eta\fud{a}{m}f^m-e_m
\chi^{ma}+\eta_m \Omega^{ma}\,,\\ res_{02} &= - [ \chi^{a[2]} R^{am}
f_m - \Omega^{a[2]} R^{am} \xi_m ] H_{a[3]} \label{res02}\,.
\end{align}

The only possibility to compensate these terms is to consider
higher-derivative non-minimal interactions and their $AdS$
deformations.

\paragraph{Cubic vertices with four derivatives: Vertex $\Omega^{a[3]}\Omega^{a[3]} R$.} The most general ansatz
for this vertex has the form
\begin{eqnarray}
{\cal L}_{41} &=& [ a_1 \Omega^{a[2]}{}_m \Omega^{a[2]}{}_m R^{m[2]}
+ a_2 \Omega^{a[3]} \Omega^a{}_{m[2]} R^{m[2]} +
\nonumber \\
 && + a_3 \Omega^{a[2]}{}_m \Omega^{am[2]} R^a{}_m +
a_4 \Omega^{am[2]} \Omega^a{}_{m[2]} R^{a[2]} ] H_{a[4]}
\label{OmegaAAAfour}\,.
\end{eqnarray}
However, due to the identity
$$
0 = \Omega^{a[3]} \Omega^{a[2]}{}_m h_m R^{m[2]} H_{a[5]} \sim [ - 3
\Omega^{a[2]}{}_m \Omega^{a[2]}{}_m - 2 \Omega^{a[3]}
\Omega^a{}_{m[2]} ] R^{m[2]} H_{a[4]}\,,
$$
the first and second terms are not independent. Let us put $a_2 =
0$. Then using the on-shell identities $h_m h_m R^{m[2]a} = 0$ and
$h_m R^{ma} = 0$, the $\chi^{a[3]}$ variation of the action can be
casted into the form
$$
\delta_0{\cal L}_{41}=[ \frac{(2a_1+a_4)}{3} \chi^a{}_{m[2]}
R^{a[3]} R^{m[2]} + 2(a_3-2a_4) \chi^{am[2]} R^{a[2]}{}_m R^a{}_m ]
H_{a[4]}\,.
$$
This enforces $a_3 = 2a_4$, while the first term can be compensated
by
\begin{equation}
\delta_1 \Phi^a = - \frac{(2a_1+a_4)}{3} \chi^a{}_{m[2]} R^{m[2]}\,.
\label{Yu421}
\end{equation}
A few comments are in order.
\begin{itemize}
\item As we see this vertex does not deform the gauge algebra. It may
seem that we took too many derivatives, but we were not able to
avoid this four-derivative vertex.
\item In all subsequent calculations it is crucial that the
combination $(2a_1+a_4)$ is non zero. Maybe the relation between
these two parameters becomes clear in the first order formalism, but
for simplicity in what follows we put $a_4=0$.
\end{itemize}

At this stage we have to consider the $\AdS$-covariantization of
this vertex, taking into account that now $D^2\neq0$. There are two
sources for non-invariance of cubic vertices in this case: terms
proportional to $m$ in the definition of the curvature tensors and
in the gauge transformations
$$
\Delta R^{a[3]} =  \frac{m}{3} h^a \Omega^{a[2]}, \qquad \delta_1
\Omega^{a[3]} = \frac{m}{3} h^a \chi^{a[2]}
$$
and this produces
\begin{equation}
res_{11} = \frac{2ma_1}{3} [ - \chi^{a[2]}{}_m ( h_m \Omega^{a[2]} +
2 h^a \Omega^a{}_m ) + \Omega^{a[2]}{}_m ( h_m \chi^{a[2]} + 2 h^a
\chi^a{}_m ) ] R^{m[2]} H_{a[4]}\,. \label{res11}
\end{equation}

\paragraph{Cubic vertices with four derivatives: Vertex $\Omega^{a[2]}\Omega^{a[2]} R$.} As far as we know, the
only possible vertex with four derivatives and bilinear in the hook
sector looks like
\begin{equation}
{\cal L} = f^a K^{a[2]} R^{a[2]} H_{a[5]}\,. \label{Yuabove423}
\end{equation}
It is similar to the first nontrivial term in the decomposition of
the Gauss--Bonnet invariant but now for two different spin-$2$
fields. By construction such a vertex exists in $d \geqslant 5$
only. For the modified $\frac32$-formalism it is convenient to
rewrite this vertex in the form $\Omega\Omega R$. So in what follows
we will use
\begin{equation}
{\cal L}_{42} = a_2 [ \Omega^{am} \Omega^a{}_m R^{a[2]} -
\Omega^{a[2]} \Omega^a{}_m R^{am} ] H_{a[4]}\label{OmegaAAfour}\,.
\end{equation}
As for the $AdS$-covariantization of this vertex, we again have two
sources for non-invariance --- terms proportional to $m$ in the
curvature and gauge transformations
$$
\Delta K^{a[2]} = - m h_m \Omega^{ma[2]}, \qquad \delta_1
\Omega^{a[2]} = - m h_m \chi^{ma[2]}\,.
$$
This produces
\begin{eqnarray}
res_{12} &=& - ma_2 [ 2 \Omega^a{}_m h_m \chi^{m[2]a} R^{a[2]} +
\Omega^{a[2]} h_m \chi^{am[2]} R^a{}_m - h_m \chi^{ma[2]}
\Omega^{an}
R^a{}_n ] H_{a[4]} + \nonumber \\
 && + ma_2 [ 2 \chi^a{}_m h_m \Omega^{m[2]a} R^{a[2]} + \chi^{a[2]}
h_m \Omega^{am[2]} R^a{}_m - h_m \Omega^{ma[2]} \chi^{an} R^a{}_n ]
H_{a[4]} \label{res12}\,.
\end{eqnarray}

\paragraph{Vertices with three derivatives.}

As we have seen from previous subsection, the four-derivative
vertices produce contributions to the $\chi^{a[3]}$- and
$\chi^{a[2]}$-variations only. It means that any variation under the
$z^a$- and $\xi^a$-transformations for the three derivatives vertex
has to be compensated by corrections to gauge transformations only.
This put severe restrictions on such vertices. The only one we have
managed to find is
\begin{equation}
{\cal L}_3 = b_0 \Omega^{a[3]} f_m R^{am} H_{a[4]}, \qquad \delta
\Phi^a = - b_0 R^{am} \xi_m\,. \label{threederivvertex}
\end{equation}
So we have only one new parameter $b_0$ to compensate for the
non-invariance coming from the minimal interactions (\ref{res01})
and (\ref{res02}) on the one hand and the non-invariance coming from
four derivatives vertices (\ref{res11}) and (\ref{res12}) on the
other hand. Happily, with a heavy use of on-shell relations $h_m h_m
R^{am[2]} = 0$ and $h_m K^{am} = 0$ one can show that all residual
variations can be canceled provided we set
\begin{equation}
a_1 = - \frac{3}{4(d-3)m^2}, \qquad a_2 = - \frac{3}{4m^2}, \qquad
b_0 = \frac{1}{m}
\end{equation}
and introduce important corrections to the gauge transformations
\begin{equation}
\delta_1 e^a = \frac{3}{2(d-3)m} [ \chi^{am[2]} \Omega_{m[2]} -
\Omega^{am[2]} \chi_{m[2]} ]\,.
\end{equation}
Thus we finally obtain a non-trivial deformation of the gauge
algebra. To summarize, we found the following cubic vertex and
corresponding gauge transformations
\begin{align}
{\cal L}_1 &= \left[ \frac32\, e_m h_m \Omega^{m[2]a} \Omega^{a[3]}
-\frac32\, e^a h_m \Omega^{m[2]a} \Omega_m{}^{a[2]} - \Omega^{a[3]}
\omega^a{}_m
\Phi^m \right] H_{a[4]}+\frac{1}{m}\, \Omega^{a[3]} f_m R^{am} H_{a[4]}\nonumber\\
& - \frac{3}{4(d-3)m^2} \,\Omega^{a[2]}{}_m \Omega^{a[2]}{}_m
R^{m[2]}H_{a[4]} - \frac{3}{4m^2}\left[ \Omega^{am} \Omega^a{}_m
R^{a[2]} - \Omega^{a[2]} \Omega^a{}_m R^{am} \right] H_{a[4]}\,,
\end{align}
\begin{align}
\delta_1 \Phi^a &=\eta^{am} \Phi_m + \eta_m h_m \Omega^{m[2]a}- e_m
h_m \chi^{m[2]a} - \omega^{am} z_m+\frac{1}{2(d-3)m^2} \chi^a{}_{m[2]} R^{m[2]}\,, \\
\delta_1 f^a &=-\omega\fud{a}{m} \xi^m +\eta\fud{a}{m}f^m-e_m
\chi^{ma}+\eta_m \Omega^{ma}\,,\\ \delta_1 e^a &= \frac{3}{2(d-3)m}
[ \chi^{am[2]} \Omega_{m[2]} -
\Omega^{am[2]} \chi_{m[2]} ]\,.
\end{align}

\subsection{St\"uckelberg formulation and Fradkin--Vasiliev approach}
\label{sec:simplehookSTFV}

In this section we consider application of the Fradkin--Vasiliev
approach to the St\"uckelberg description of the hook field. First
of all we have to rewrite the free Lagrangian in terms of gauge
invariant curvatures. The result reads
\begin{equation}
{\cal L}_0 = [ a_1 R^{a[2]m} R^{a[2]}{}_m + a_2 K^{a[2]} K^{a[2]} +
a_3 R^{a[3]} K^a ]\, H_{a[4]}\,,
\end{equation}
where
\begin{equation}
\frac{(d-4)a_1}{3} + a_2 = - \frac{3}{4m^2}, \qquad a_3 = -
\frac{1}{m}\,.
\end{equation}
Again we see that there is an ambiguity in the choice of
coefficients but the choice will be fixed after switching on
interactions.

Now we have to construct deformed curvatures both for the hook field
and for the graviton, so that the variations will be proportional to
the free curvatures. For the graviton the result is easy to find
\begin{eqnarray}
\hat{R}^{a[2]} &=& R^{a[2]} + ma_0 [ \Omega^{am[2]}
\Omega^a{}_{m[2]}
- \frac{2}{3} \Omega^{am} \Omega^a{}_m ]\,, \\
\hat{T}^a &=& T^a + a_0 \Omega^{am[2]} \Omega_{m[2]}\,,
\end{eqnarray}
with the corresponding variations having the form
\begin{eqnarray}
\delta \hat{R}^{a[2]} &=& ma_0 [ R^{am[2]} \chi^a{}_{m[2]} -
\frac{2}{3} K^{am} \chi^a{}_m ]\,, \\
\delta \hat{T}^a &=& a_0 [ - \chi^{am[2]} K_{m[2]} + R^{am[2]}
\chi_{m[2]} ]\,.
\end{eqnarray}

The deformations for the hook's curvatures simply correspond to the
standard Lorentz minimal coupling
\begin{eqnarray}
\hat{R}^a &=& R^a + \omega^{am} \Phi_m - h_m e_m \Omega^{m[2]a} -
\frac{m}{3} ( e^a h_m + h^a e_m) f^m\,, \\
\hat{R}^{a[3]} &=& R^{a[3]} + \omega^a{}_m \Omega^{ma[2]} +
\frac{m}{3} e^a \Omega^{a[2]}\,, \\
\hat{K}^a &=& K^a + \omega^{am} f_m + e_m \Omega^{ma}\,, \\
\hat{K}^{a[2]} &=& K^{a[2]} + \omega^a{}_m \Omega^{ma} - m e_m
\Omega^{ma[2]}\,.
\end{eqnarray}

In what follows we will need only the part of the variation that
does not vanish on-shell. It has a simple form
\begin{equation}
\delta \hat{R}^{a[3]} = R^a{}_m \chi^{ma[2]}\,,  \qquad \delta
\hat{K}^{a[2]} = R^a{}_m \chi^{ma}\,,  \qquad \delta \hat{K}^a =
R^{am} \xi_m\,.
\end{equation}

Now, following the general procedure, we consider the interacting
Lagrangian
\begin{equation}
{\cal L}_0+ {\cal L}_1= [ a_1 \hat{R}^{a[2]m} \hat{R}^{a[2]}{}_m +
a_2 \hat{K}^{a[2]} \hat{K}^{a[2]} + a_3 \hat{R}^{a[3]} \hat{K}^a +
\frac{1}{4\lambda^2} \hat{R}^{a[2]} \hat{R}^{a[2]} ] H_{a[4]}\,.
\end{equation}
The next problem is to adjust the coefficients so that all
variations vanish on-shell. For the $\chi^{a[3]}$-transformations we
obtain\footnote{The deformed curvatures introduced above are
associated with corresponding $\delta^g_1$, see Section
\ref{sec:CubicInteractions}.}
$$
\delta^g_1{\cal L}_0+ \delta_0{\cal L}_1=[ - 4a_1 R^{a[2]m} R^{am}
\chi^a{}_{m[2]} + \frac{3a_0}{m} R^{am[2]} R^{a[2]} \chi^a{}_{m[2]}
] H_{a[4]}\,.
$$
Using the on-shell relations $h_m h_m R^{am[2]} = 0$ and $h_m R^{am}
= 0$ one can show that the following identity holds
$$
[ - 2 R^{a[2]m} R^{am} \chi^a{}_{m[2]} + R^{am[2]} R^{a[2]}
\chi^a{}_{m[2]} ] H_{a[4]} = 0\,.
$$
Thus we have to put
\begin{equation}
a_1 = \frac{3a_0}{2m}\,.
\end{equation}
At the same time, for the $\chi^{a[2]}$-transformations we obtain
$$
\delta^g_1{\cal L}_0+ \delta_0{\cal L}_1=[ 4a_2 K^{a[2]} R^{am}
\chi^a{}_m + \frac{2a_0}{m} K^{am} R^{a[2]} \chi^a{}_m ] H_{a[4]}\,.
$$
Again, using the on-shell relations $h_m K^{am} = 0$ and $h_m R^{am}
= 0$, one can show that
$$
[ K^{a[2]} R^{am} \chi^a{}_m + K^{am} R^{a[2]} \chi^a{}_m ] H_{a[4]}
= 0\,.
$$
Therefore, we set
\begin{equation}
a_2 = \frac{a_0}{2m} = \frac{a_1}{3}\,.
\end{equation}
In particular, the last relation fixes the ambiguity in the free
Lagrangian giving us
\begin{equation}
a_1 = - \frac{9}{4(d-3)m^2} \quad \Longrightarrow \quad a_0 = -
\frac{3}{2(d-3)m}\,.
\end{equation}

\paragraph{Going from St\"uckelberg to ASV.} We have already mentioned that the St\"uckelberg formulation is related to
the ASV one through the partial gauge fixing, see (\ref{gaugetrA})
and below. It is instructive to see how this procedure works in the
interacting case. First of all, using the fact that for any non-zero
$\lambda$ we have $\delta f^a \sim z^a$, we can choose the gauge
$f^a = 0$. Then, the corresponding torsion equation
\begin{equation}
\hat{K}^a = h_m \Omega^{ma} - m \Phi^a + e_m \Omega^{ma} = 0
\end{equation}
gives us
\begin{equation}
\Phi^a = \frac{1}{m} ( h_m + e_m) \Omega^{ma}\,.
\end{equation}
As a consequence, the second torsion equation $R^a = 0$ does not
carry any new information, leaving us with two non-trivial
curvatures only
\begin{eqnarray}
\hat{R}^{a[3]} &=& D \Omega^{a[3]} + \omega^a{}_m \Omega^{ma[2]} +
\lambda^2 (h^a + e^a) \Omega^{a[2]}\,, \\
\hat{K}^{a[2]} &=& D \Omega^{a[2]} + \omega^a{}_m \Omega^{ma} - (h_m
+ e_m) \Omega^{ma[2]}\,.
\end{eqnarray}
Here we have made the rescaling $\Omega^{a[2]} \to m\Omega^{a[2]}$
and $K^{a[2]} \to m K^{a[2]}$ in accordance with the fact that
$\Omega^{a[2]}$ plays now the role of a physical field. After such a
rescaling the deformed Riemann tensor has the form
\begin{equation}
\hat{R}^{a[2]} = R^{a[2]} + ma_0 [ \Omega^{am[2]} \Omega^a{}_{m[2]}
- 2 \lambda^2 \Omega^{am} \Omega^a{}_m ]\,,
\end{equation}
while the interacting Lagrangian can be written as follows
\begin{equation}
{\cal L} = a_1 [ \hat{R}^{a[2]m} \hat{R}^{a[2]}{}_m + \lambda^2
\hat{K}^{a[2]} \hat{K}^{a[2]} ] H_{a[4]} + \frac{1}{4\lambda^2}
\hat{R}^{a[2]} \hat{R}^{a[2]} H_{a[4]}\,,
\end{equation}
which is to be compared with the genuine ASV action
(\ref{dynactionAnsatz}).

\paragraph{Flat limit.} According to the general analysis of flat
limit of higher-spin cubic actions in \AdS{} done in
\cite{Boulanger:2008tg}, one can always rescale the fields and
dimensionful coupling constants in such a way that the flat limit of
the \AdS{} action retains only the quadratic kinetic terms and the
cubic highest derivative terms. Of course, the \AdS{}-covariant
derivatives are replaced by the flat partial ones. In our case the
cubic vertex that will survive in the flat limit is the
four-derivative one given by (\ref{OmegaAAAfour}),
(\ref{Yuabove423}) that are abelian. Note that there is always a
freedom in adding total derivative terms and making
field-redefinitions, c.f. (\ref{Yuabove423}) and
(\ref{OmegaAAfour}).

\subsection{ASV formulation and Fradkin--Vasiliev approach}
\label{sec:simplehookFV}

In this Section we would like to test gravitational interactions for
the simplest case of spin-$[2,1]$ gauge fields, i.e. we are
interested in $[2,1]-[2,1]-[1,1]$ cubic vertices. The simplicity is
due to the fact that the spin-$[2,1]$ field is described by a
one-form.

We introduce the following set of one-form gauge fields
$\{e^a,\omega^{ab},\Omega^{a[2]},\Omega^{a[3]}\}$ where
$\{e^a,\omega^{ab}\}$ are the dynamical one-form gauge fields in the
spin-2 sector. As recalled in Section \ref{sec:typepq}, the two
fields $\{\Omega^{a[2]},\Omega^{a[3]}\}$ correspond to the one-forms
needed to describe an irreducible and unitary $[2,1]$-type gauge
field in $\AdS$ within the ASV formulation.

Quadratic corrections to curvatures are made by replacing background
tetrad $h^a$ and Lorentz spin-connection $\varpi^{a,b}$ with
$h^a+e^a$ and $\varpi^{a,b}+\omega^{a,b}$, respectively. Quadratic
contributions to the torsion and Riemann curvature are determined
from the most general Ansatz by requiring curvatures to be gauge
invariant up to order $g$. Denoting the total vielbein
$\mathbf{e}^a=h^m+e^m\,$, the result is: %
\begin{eqnarray}
 T^{a} &=& D e^a + \mathbf{e}_b\,\omega^{ba} - g\,\Omega^{ab[2]}\,\Omega_{b[2]}\quad, \\
 R^{ab} &=& D \omega^{ab} + \omega^{a}{}_c\,\omega^{cb} - \Lambda \,\mathbf{e}^a \,\mathbf{e}^b
 - 2\,\Lambda \, g \,\Omega^{ac}\,\Omega^b{}_{c} - g\, \Omega^{ac[2]}\,\Omega^b{}_{c[2]}\quad,  \\
 K^{a[2]} &=& D \Omega^{a[2]} - \mathbf{e}_b\,\Omega^{ba[2]} + \omega^{a}{}_b \,\Omega^{ba}\quad, \\
 R^{a[3]} &=& D \Omega^{a[3]} +  \Lambda \mathbf{e}^a \,{\Omega}^{a[2]} +
 \omega^{a}{}_b\,\Omega^{ba[2]} \quad .
 \end{eqnarray}
The Yang--Mills-like gauge transformation are
\begin{eqnarray}
 \delta e^{a} &=&  {\rm d} \xi^a + {\xi}_b\,\omega^{ab} - {e}_b\,\xi^{ab} + g\,\eta^{ab[2]}\,\Omega_{b[2]}
               - g\, \Omega^{ab[2]}\,\eta_{b[2]} \quad,
     \\
 \delta \omega^{ab} &=& {\rm d} \xi^{ab} - \xi^{a}{}_c\,\omega^{cb} + \omega^{ac}\,\xi^{b}{}_c
              + \Lambda \,({\xi}^a \,e^b - e^a\,{\xi}^b)
              + 2\,\Lambda \, g \,\eta^{ac}\,\Omega^b{}_{c}
    \nonumber \\
              && - \;2\,\Lambda \, g \,\Omega^{ac}\,\eta^b{}_{c}
              + g\, \eta^{ac[2]}\,\Omega^b{}_{c[2]} - g\, \Omega^{ac[2]}\,\eta^b{}_{c[2]}\quad,
      \\
  \delta \Omega^{a[2]} &=& {\rm d} \eta^{a[2]} + {\xi}_b\,\Omega^{ba[2]} - {e}_b\,\eta^{ba[2]}
              - \xi^{a}{}_b \, \Omega^{ba} + \omega^{a}{}_b \,\eta^{ba} \quad,
      \\
  \delta \Omega^{a[3]} &=& {\rm d} \eta^{a[3]} - \Lambda {\xi}^a \,\Omega^{a[2]}
                + \Lambda {e}^a \,\eta^{a[2]} - \xi^{a}{}_b\,\Omega^{ba[2]}
                +  \omega^{a}{}_b\,\eta^{ba[2]}  \quad ,
\end{eqnarray}
and accordingly, for the curvatures:
\begin{eqnarray}
 \delta T^{a} &=&  {\xi}_b\,R^{ab} - {T}_b\,\xi^{ab} + g\,\eta^{ab[2]}\,K_{b[2]}
               - g\,R^{ab[2]}\,\eta_{b[2]} \quad,
\label{deltaTa}\\
 \delta R^{ab} &=& - \xi^{a}{}_c\,R^{cb} + R^{ac}\,\xi^{b}{}_c
              + \Lambda \,({\xi}^a \,T^b - T^a\,{\xi}^b)
              + 2\,\Lambda \, g \,\eta^{ac}\,K^b{}_{c}\nonumber \\
              && - \;2\,\Lambda \, g \,K^{ac}\,\eta^b{}_{c}
              + g\, \eta^{ac[2]}\,R^b{}_{c[2]} - g\, R^{ac[2]}\,\eta^b{}_{c[2]}\quad,
\label{deltaRab} \\
  \delta K^{a[2]} &=& {\xi}_b\,R^{ba[2]} - {T}_b\,\eta^{ba[2]}
              - \xi^{a}{}_b \,K^{ba} + R^{a}{}_b \,\eta^{ba} \quad,
\label{deltaTak} \\
  \delta R^{a[3]} &=& - \Lambda {\xi}^a \,K^{a[2]} +\Lambda {T}^a \,\eta^{a[2]}
                   - \xi^{a}{}_b\,R^{ba[2]} + R^{a}{}_b\,R^{ba[2]}  \quad
\label{deltaRak1}.
 \end{eqnarray}
The on-mass-shell conditions for $[2,1]$-type fields read, (\ref{onmassshellpqA})-(\ref{onmassshellpqB}),
\begin{eqnarray}
&{K}^{a[2]}_0 ~=~ h_b h_b \; {\cal C}^{a[2],b[2]} \quad,\qquad
{R}^{a[3]}_0  ~=~ h_b h_b \; {\cal C}^{a[3],b[2]} \quad ,&
\label{onshellline}
\end{eqnarray}
while the spin-2 sector gives the constraints
\begin{eqnarray}
{ T}^a_0 &=& 0\quad , \qquad { R}^{a[2]}_0 ~=~ h_b h_b \; {\cal
W}^{aa,bb} \quad. \label{onshellspin2}
\end{eqnarray}
where the linearized quantities are indicates by calligraphic symbols.
The \lorentz-tensors $\{ {\cal C}^{a[2],b[2]},{\cal C}^{a[3],b[2]},
{\cal W}^{a[2],b[2]}\}$ are irreducible tensors of symmetry type
$[2,2],[3,2]$ and $[2,2]\,$, respectively. \vspace*{.2cm}

We take the following Ansatz for the action
\begin{eqnarray}
& S[e^a,\omega^{ab},\Omega^{a[2]},\Omega^{a[3]}] = \frac{1}{2}\; \int  ( R^{uu}\wedge R^{vv} + a_1 \; K^{uu}
\wedge K\fud{vv}{}
 + a_2 \; R^{uua}\wedge R\fud{vv}{a})\wedge H_{uuvv} \,, &
 \label{dynactionAnsatz}
\end{eqnarray}
where it is understood that the quartic terms are neglected at this
order in perturbation. The variation of the above action can be
evaluated using (\ref{deltaTa})--(\ref{deltaRak1}), keeping only
terms bilinear in the fields and linear in the gauge parameter. In
other words, after taking the gauge variation inside the action, the
curvatures are replaced by their linearized expressions that are
then constrained according to (\ref{onshellline}) and
(\ref{onshellspin2}).

Denoting
\begin{align}
\boldsymbol{A}&=\int H_{u[0]} \; {\cal C}_{uu,vv} \,\eta\fud{v}{a}
\,{\cal W}^{va,uu}
\quad, \\
\boldsymbol{B}&=\int H_{u[0]} \;{\cal C}_{uu,vv}
\,\eta\fud{v}{a[2]} \,{\cal W}^{va[2],uu}
\quad, \\
\boldsymbol{C}&=\int H_{u[0]} \;{\cal W}_{a[2],mn} \,\xi_c \,{\cal
W}^{ca[2],mn} \quad,
\end{align}
the Fradkin--Vasiliev consistency condition gives the following
constraint on the free parameters entering  the action
$S[e^a,\omega^{ab},\Omega^{a[2]},\Omega^{a[3]}]$:
\begin{align}
 \left[ 2 \Lambda g - a_1\; \right] \,\boldsymbol{A}
 + \left[ g - \frac{ a_2}{2}\;\right] \,\boldsymbol{B}
 + \left[ \frac{ a_1}{2}\; - \frac{ a_2\Lambda }{2}\;\right] \,\boldsymbol{C} =0\quad.
\end{align}
This admits the solution
\begin{equation}\label{graviratio}
a_1~=~ 2 g\, \,\Lambda  \quad, \qquad a_2~=~ 2 g \quad .
\end{equation}
We see that the ratio $a_1/a_2$ is completely fixed by the
consistency of the action (\ref{dynactionAnsatz}).

Within the manifestly $\AdS$-covariant ASV formulation the
computations are basically the same, but one has to take into
account a fewer number of terms as some of them join together into
single $AdS$-covariant objects.

One important remark is that switching on gravitational interactions
dictates the relative coefficient $a_1/a_2$ in a way that manifestly
$AdS$-covariant ASV action acquires the most simple form
\begin{align}
S=\frac12\int R^{UUA}\wedge R\fud{VV}{A}\wedge H_{UUVV}\,.
\end{align}

Unfortunately, it is impossible to take a meaningful flat limit
because of discontinuity in the number of physical degrees of
freedom. Note that the two fields of ASV formulation correspond to
fields similar to Lorentz spin-connection rather than a tetrad-like
fields, which can be excluded if $\Lambda\neq0$ as explained in
section \ref{sec:typepq}. Even at the linearized level the action
for $\Omega^{aa}\fm{1}$ and $\Omega^{aaa}\fm{1}$ reduces to a
nonunitary theory because no additional gauge symmetry reappear.

\section{Gravitational interactions of type$\,$-$\boldsymbol{[2,1]}$ field in flat space:\\
metric-like St\"uckelberg formulation and BV-BRST approach}
\setcounter{equation}{0}\label{sec:flatlimit}

In section \ref{sec:simplehookSTFV} we have seen that, in the flat
limit, no nonabelian interaction could survive in the St\"uckelberg
action. In this section we discuss, in the metric-like formalism,
the flat limit of the interacting Lagrangian obtained in this paper
by addressing the related problem of determining all the possible
interactions that the hook field can have with gravity in flat
space. We show that indeed it is not possible to build nonabelian
interactions in flat space, thereby strengthening the results of the
section \ref{sec:simplehookSTFV}.

The couplings will involve the following three types of gauge
fields: a $[2,1]$-type field $T_{\mu\nu,\rho}=-T_{\nu\mu,\rho}$ (we
use the antisymmetric convention in this section), the graviton and
the St\"uckelberg companion of the hook field which is of the same
symmetry type as the graviton, \emph{i.e.} in metric-like formalism
it is a rank-two symmetric gauge field\footnote{See Appendix
\ref{sec:app} for a metric-like presentation of the free
St\"uckelberg action for a hook field in $\AdS{}$.}. There are not
many cubic terms that can consistently couple these fields in a flat
background and we will show that none of them is compatible with the
existence of a nonabelian gauge algebra at the first nontrivial
order, in sharp contrast to what happens for totally-symmetric gauge
fields. It means that it is definitely the cosmological constant
that is responsible for the nonabelian nature of the interactions we
have presented in this paper.

In flat space, the problem of the self-interactions for arbitrary
type-$[p,q]$ gauge fields was thoroughly studied in
\cite{Bekaert:2002uh,Boulanger:2004rx,Bekaert:2004dz} via the
cohomological reformulation \cite{Barnich:1993vg} of the N\"other
procedure for constructing consistent interactions
\cite{Berends:1984rq}. For any $[p,q]$-type gauge fields in flat
space, all the relevant cohomological groups have been computed in
\cite{Bekaert:2002uh,Boulanger:2004rx,Bekaert:2004dz} to which we
refer for more details.

Without entering too much into the details of the antifield
formulation for $[p,q]$-type gauge fields in flat space
\cite{Boulanger:2004rx,Bekaert:2004dz}, we will give here a list of
the various possible cubic couplings between a $[2,1]$-type gauge
field and a set of two different gravitons, the physical graviton
and the St\"uckelberg companion of the $[2,1]$-type gauge field. We
will show that there is no way to build nonabelian cubic vertices
among the three species of fields considered here if there is at
least one hook field occurring in the vertex. In the case of the
cubic coupling between colored gravitons, it is a result of
\cite{Boulanger:2000rq} that there is no nontrivial nonabelian
interactions mixing colored gravitons. Therefore, in the case of
interactions between the St\"uckelberg field and the physical
graviton, there is no possibility for nonabelian interactions.

\paragraph*{Sector of the mixed-symmetry gauge field}

The spectrum of fields and antifields in the sector of the
$[2,1]$-type gauge field is given by
\begin{itemize}
\item the fields $T_{\alpha\beta,\gamma}$ with ghost number ($gh$) zero and
antifield number ($antigh$) zero;
\item the ghosts $S_{\alpha\beta}=S_{(\alpha\beta)}$ and $A_{\alpha\beta}=A_{[\alpha\beta]}$
with $gh=1$ and $antigh = 0$;
\item the ghosts of ghosts $B_{\alpha}$ with $gh = 2$ and $antigh = 0$,
which appear because of the reducibility relations;
\item the antifields $T^{* [\alpha\beta]\gamma}$, with ghost number minus one
$gh = -1$ and $antigh = 1$;
\item the antifields $S^{*(\alpha\beta)}$ and $A^{*[\alpha\beta]}$ with $gh = -2$
and $antigh = 2$;
\item the antifields $B^{*\alpha}$ with $gh = -3$ and $antigh = 3\,$.
\end{itemize}
\vspace*{.3cm}

\noindent Note that the antifield number is sometimes also called
``antighost number''. The BRST differential for the free theory
takes the simple form
\begin{equation}
 s = \delta + \gamma\quad .
\end{equation}
A grading is associated with each of these differentials : $\gamma$
increases by one unit the ``pure ghost number" denoted {\it{puregh}}
while the Koszul--Tate differential $\delta$ increases the antighost
number {\it{antigh}} by one unit. The ghost number {\it{gh}} is
defined by
\begin{equation}
{\it{gh}}={\it{puregh}}-{\it{antigh}}.
\end{equation}
The action of the differentials $\gamma$ and $\delta$ on all the
fields of the formalism is displayed in Table \ref{table1} that
indicates also the  pureghost number, antighost number, ghost number
and grassmannian parity of the various fields. \vspace{3mm}

\begin{table}[h]
\begin{tabular}{|c|c|c|c|c|c|c|}
\hline Z & $\gamma(Z)$  & $\delta(Z)$  & $puregh(Z)$  & $antigh(Z)$
& $gh(Z)$  & parity \\ \hline
$T_{\alpha\beta,\gamma}$  &  $\gamma T_{\alpha\beta,\gamma}$  & $0$  &$0$  & $0$  &$0$ &$0$ \\
$S_{\alpha\beta}$ & $6\partial_{(\alpha}B_{\beta)}$ & $0$ & $1$ & $0$ & $1$ & $1$ \\
$A_{\alpha\beta}$ & $2\partial_{[\alpha}B_{\beta]}$ & $0$ & $1$ & $0$ & $1$ & $1$ \\
$B_{\alpha}$       & $0$ &  $0$      & $2$  & $0$ & $2$ &  $0$ \\
$T^*_{\beta_1\beta_2,}{}^{\alpha_1}$ & $0$ &
$ -16\,\delta^{\alpha[4]}_{\beta[4]} \,K_{\alpha[3],}{}^{\beta[2]}$
& $0$
& $1$ & $-1$ & $1$ \\
$S^{*\alpha\beta}$ & $0$ & $
-2\,\partial_{\gamma}T^{*\gamma(\alpha,\beta)}$ & $0$ & $2$ & $-2$ &
$0$
\\
$A^{*\alpha\beta}$ & $0$ &
$-6\,\partial_{\gamma}T^{*\gamma[\alpha,\beta]}$
& $0$ & $2$ & $-2$ & $0$ \\
$B^{*\alpha}$ & $0$ &
$6\,\partial_{\mu}S^{*\mu\alpha}+2\,\partial_{\mu}A^{*\mu\alpha}$ &
$0$& $3$ & $-3$ & $1$
\\
\hline
\end{tabular}
\caption{Action of the differentials $\gamma$ and $\delta$ in the
sector of the $[2,1]$-types gauge field, where $\gamma
T_{[\alpha\beta]\gamma}=2(\partial_{[\alpha}S_{\beta]\gamma}+\partial_{[\alpha}A_{\beta]\gamma}
-\partial_{\gamma}A_{\alpha\beta})\,$,
$K_{\alpha[3],}{}^{\beta[2]}=-\frac{3}{4}\;\partial^{[\beta_1}\partial_{[\alpha_1}
T_{\alpha_2\alpha_3],}{}^{\beta_2]}$ is the gauge-invariant
linearized curvature tensor and $\delta^{\alpha[4]}_{\beta[4]} =
\delta^{\alpha_1}_{[\beta_1}\delta^{\alpha_2}_{\beta_2}
\delta^{\alpha_3}_{\beta_3}\delta^{\alpha_4}_{\beta_4]}$.}
\label{table1}
\end{table}

It is convenient to perform a change of variables in the $antigh=2$
sector in order for the Koszul-Tate differential to take a simpler
expression when applied on all the antifields of {\it{antigh}}
$\geqslant$ $2\,$. We define
\begin{equation}
 C^{*\alpha\beta}=3\,S^{*\alpha\beta} + A^{*\alpha\beta}.
\end{equation}
It leads to the following simple expressions
\begin{eqnarray}
\delta C^{*\alpha\beta}&=&-6 \,\partial_{\gamma}
T^{*[\gamma\alpha]\beta},
\\
\delta B^{*\mu}&=&2\,\partial_{\nu}C^{*\nu\mu}.
\end{eqnarray}

Following the results of \cite{Bekaert:2002uh,Bekaert:2004dz}, in
order to perturbatively deform the solution $W^{(0)}$ of the master
equation for the free theory $(W^{(0)}W^{(0)})_{A.B.}=0$ [where
$(\cdot,\cdot)_{A.B.}$ denotes the antibracket]  into
$W=W^{(0)}+g\,W^{(1)}+\ldots$ with $W^{(1)} = \int
d^dx\;(a_0+a_1+a_2+a_3)\,$ where $gh(a_i)=0\,$ and
$antigh(a_i)=i\,$, one has to solve the cocycle equation
$s\,a+d\,c=0$ where  $a=a_0+a_1+a_2+a_3\,$ and $c=c_2+c_1+c_0\,$.
The descent of equations coming from the decomposition of $sa+dc=0$
with respect to the antifield number is
\begin{eqnarray}
 \delta a_1 + \gamma a_0 + dc_0 &=& 0 \quad, \label{descent1}\\
\delta a_2 + \gamma a_1 + dc_1 &=& 0  \quad,\label{descent2}\\
\delta a_3 + \gamma a_2 + dc_2 &=& 0  \quad,\label{descent3}\\
                    \gamma a_3 &=& 0 \label{descent4}\quad.
\end{eqnarray}
In order to solve this system, one starts with $a_3$ that must
belong to the cohomological group $H(\gamma)$ and plug it into the
equation $\delta a_3 + \gamma a_2 + dc_2 = 0$ that must be solved
for $a_2\,$. If that is possible, one has thereby ``lifted'' or
``integrated'' $a_3$ to $a_2\,$. In case such an integration is not
obstructed, one plugs $a_2$ into the next equation $\delta a_2 +
\gamma a_1 + dc_1 = 0$ and try to solve it for $a_1\,$. If $a_2$ can
be lifted to an $a_1\,$, one finally try to solve $\delta a_1 +
\gamma a_2 + dc_0 = 0$ for $a_0$ which is the vertex appearing in
the deformed Lagrangian. The deformations of the gauge algebra
appear into $a_2$ while $a_1$ contains the deformations of the gauge
transformations. The element $a_3$ contains information about the
deformation of the reducibility transformations. \vspace*{3mm}

Important ingredients for the construction of the various elements
$\{a_i\}\,$, $i=0,1,2,3\,$, are the following cohomological groups:
\begin{enumerate}
 \item[(i)] $H(\gamma)\,$, the cohomology of $\gamma\,$, is isomorphic to the algebra
\\ $\left\{ f \left( [K_{\alpha_1\alpha_2\alpha_3,\beta_1\beta_2}],[\Phi^*],B_{\mu},
H^A_{\alpha_1\alpha_2\alpha_3}\right) \right\}$ of functions of the
generators, where $H^A_{\alpha_1\alpha_2\alpha_3} =
\partial_{[\alpha_1}A_{\alpha_2\alpha_3]}\,$, $[\Phi^*]$ denotes
collectively all the antifields and their derivatives and similarly
$[K_{\alpha_1\alpha_2\alpha_3,\beta_1\beta_2}]$ denotes the
curvature tensor and all its derivatives;
\item[(ii)] The cohomology groups $H^d_q(\delta \vert d)$
 vanish in antifield  number $q$ strictly greater than three:
 $H^d_q(\delta \vert d) \cong 0 \, \hbox{ for } q>3\,$;
\item[(iii)] A complete set of representatives of
 $H^d_3(\delta \vert d)$ is given by the antifields $B^{*\mu}$ conjugate
 to the ghost of ghosts $B_\mu\,$, {\it{i.e.}}, $ \delta a^d_3+d a^{d-1}_{2}=0
 \Rightarrow a^d_3=\lambda_{\mu}B^{*\mu} dx^0\wedge dx^1\wedge\ldots\wedge
 dx^{d-1}+\delta b_4^d+d b_3^{d-1} $ where the $\lambda_{\mu}$'s are
 constants;
\item[(iv)]
The cohomological group $H^d_2(\delta \vert d)$ vanishes if one
considers cochains $a$ that have no explicit $x$-dependence (as it
is necessary for constructing Poincar\'e-invariant Lagrangians).
\end{enumerate}

\paragraph*{Sector of the graviton}

In the sector of the graviton fields, the relevant cohomological
analysis was performed in \cite{Boulanger:2000rq}. On top of the
spin-2 gauge field $h_{\mu\nu}\,$, the BRST--BV spectrum includes
ghost $C_\mu$ associated with the linearized diffeomorphisms
together with the antifields $h^{*\mu\nu}$ and $C^{*\mu}\,$. We have
summarized the action of the various relevant differentials on these
fields in Table~\ref{table2}.

\begin{table}[h]
\begin{tabular}{|c|c|c|c|c|c|c|}
\hline Z & $\gamma(Z)$  & $\delta(Z)$  & $puregh(Z)$  & $antigh(Z)$  & $gh(Z)$  & parity \\
\hline
$h_{\mu\nu}$  &  $2\,\partial_{(\mu}C_{\nu)}$  & $0$  &$0$  & $0$  &$0$ &$0$ \\
$C_{\mu}$ & $0$ & $0$ & $1$ & $0$ & $1$ & $1$ \\
$h^{*\nu}{}_{\mu}$ & $0$ & $ -\frac{9}{2}\;\delta^{\mu[2]}_{\nu[3]}
\,K_{\mu[3],}{}^{\nu[2]}$
& $0$ & $1$ & $-1$ & $1$ \\
$C^{*\mu}$ & $0$ & $-\partial_{\nu}h^{*\mu\nu}$ & $0$ & $2$ & $-2$ & $0$ \\
\hline
\end{tabular}
\caption{Action of the differentials $\gamma$ and $\delta$ in the
sector of the spin-2 field, where
$K_{\alpha[2],}{}^{\beta[2]}=\frac{4}{3}\;\partial^{[\beta_1}\partial_{[\alpha_1}
h_{\alpha_2}{}^{\beta_2]}$ is the gauge-invariant linearized
curvature tensor and $\delta^{\alpha[3]}_{\beta[3]} =
\delta^{\alpha_1}_{[\beta_1}\delta^{\alpha_2}_{\beta_2}
\delta^{\alpha_3}_{\beta_3]}$.} \label{table2}
\end{table}

In this sector, the relevant cohomological groups are:

\begin{enumerate}
 \item[(i)] $H(\gamma)\,$, the cohomology of $\gamma\,$, that is isomorphic to the algebra
\\ $\left\{ f \left( [K_{\alpha_1\alpha_2,\beta_1\beta_2}],[\Phi^*],
C_{\mu},\partial_{[\mu}C_{\nu]}\right)\right\}$ of functions of the
generators;
\item[(ii)] $H^d_q(\delta \vert d) \cong 0 \, \hbox{ for } q>2\,$;
\item[(iii)]
 $H^d_2(\delta \vert d)$ given by the antifields $C^{*\mu}$ conjugate
 to the ghosts $C_\mu\,$, {\it{i.e.}}, $ \delta a^d_2+d a^{d-1}_{1}=0
 \Rightarrow a^d_2=\lambda_{\mu}C^{*\mu} dx^0\wedge dx^1\wedge\ldots\wedge
 dx^{d-1}+\delta b_3^d+d b_2^{d-1} $ where the $\lambda_{\mu}$'s are
 constants.
\end{enumerate}

\paragraph*{Couplings [1,1]--[1,1]--[1,1]}

As we said above, there is no nonabelian vertices mixing
nontrivially several kinds of gravitons, therefore the only
couplings we can introduce are the Born--Infeld coupling with six
derivatives (that does not appear in the analysis in \AdS{}), and
the following four-derivative coupling in $d\geqslant 5\,$:
\begin{equation}
a_0 \sim
\delta^{b[5]}_{a[5]} \;h_b{}^{a}\,
K_{b[2]}{}^{a[2]}\,K_{b[2]}{}^{a[2]}
\end{equation}
that does not modifies the gauge transformations and is invariant up to a total derivative.
In other words, in (\ref{descent1}) there is no corresponding $a_1$ but a nonzero $c_0\,$.
As we see, this vertex brings in four derivatives and it contributes via (\ref{Yuabove423})
to making the \AdS{} nonabelian vertex in the St\"uckelberg formulation.

\paragraph*{Couplings [1,1]--[1,1]--[2,1]}

There is a candidate deformation with five derivatives between the
hook field $T_{\mu[2],\nu}$ and two gravitons $h^i_{\mu\nu}\,$,
$i=1,2\,$ (in the following we will omit the extra internal index
$i$ for simplicity of notation):
\begin{equation}
 a_0 = K_{\nu[3],}{}^{\mu[2]}\;K_{\nu[2],}{}^{\mu[2]}\,\partial^\mu h^{\mu}{}_\nu\;
\epsilon_{\mu[6]}\,\epsilon^{\nu[6]}
\end{equation}
where we recall that $K_{\nu[3],}{}^{\mu[2]}$ is the curvature
tensor for $T_{\mu[2],\nu}$ and $K_{\nu[2],}{}^{\mu[2]}$ is the
linearized curvature tensor in the spin-2 sector. It is easy to
check that this vertex is gauge invariant under linearized
transformations, up to a total derivative.

With three derivatives involved, it can be seen that there is only
one candidate associated with the following element of the
cohomology of $\gamma$: $a_1 =
T^{*\mu[2],\nu}K_{\mu[2],\nu[2]}C^\nu\,$, where $C^\nu$ is the ghost
associated with the linearized diffeomorphisms. The element $a_1\in
H(\gamma)$ encodes the information concerning a deformation of the
gauge transformations for the hook field; a deformation that does
not modify the gauge algebra which therefore remains abelian.
Explicitly, it corresponds to the transformation
\begin{equation}
 \delta^{(1)}T_{\mu_1\mu_2,\nu_1} = K_{\mu_1\mu_2,\nu_1\nu_2}\;\xi^{\nu_2} \quad .
\end{equation}
To see whether this deformations of the gauge transformations can be
integrated to a cubic vertex $a_0\,$, one has to solve the equation
$\delta a_1 + \gamma a_0 = dc_0 $ where $\delta$ is the Koszul-Tate
differential and $\gamma$ is the differential that implement the
gauge transformations. When computing $\delta a_1$, it is possible,
up to total derivatives, to make appear a $\gamma$-exact term
$\tilde a_0 = \delta^{a[4]}_{b[4]}\,\partial_{a}\Phi_{aa|}{}^b
\,h^{bc}\, K_{ac|}{}^{bb}$, but there remains a term that cannot be
written as $\gamma$-exact term up to total derivative so that in flat spacetime,
this vertex is not consistent.

Actually, looking at the classification found in
\cite{Metsaev:2005ar}, one sees that there is indeed only one
vertex, bringing five derivatives, whereas in \AdS{}, this
three-derivative vertex plays an important role in making the nonabelian
interactions we presented above, see (\ref{threederivvertex}).

\paragraph*{Couplings [1,1]--[2,1]--[2,1]}

In terms of the quantities $j_\sigma^{(a)}$, $a=1,2,3$ and $\sigma
=1,2$ relevant for the formula (8.66) of Metsaev in \cite{Metsaev:2005ar}
(that formula is reproduced below in \ref{Ruslan})),
the coupling $[1,1]-[2,1]-[2,1]$ corresponds to either
$(1,0)-(\frac{3}{2}\,,\frac{1}{2}\,)-(\frac{3}{2}\,,\frac{1}{2}\,)\,$,
$(1,0)-(\frac{3}{2}\,,\frac{1}{2}\,)-(\frac{1}{2}\,,\frac{3}{2}\,)$
or $(1,0)-(\frac{1}{2}\,,\frac{3}{2}\,)-(\frac{1}{2}\,,\frac{3}{2}\,)$
since in $d=6$ the hook field receives two Gelfand--Zetlin labels
$(s_1,s_2)$ with $|s_2|\leqslant s_1\,$,
so that the hook field can have $s_2=1$ or $s_2=-1\,$,
depending on its being self dual or anti-self dual.\footnote{See the discussion at the
beginning of Section 8.2 in \cite{Metsaev:2005ar}. Note that the restriction on
$k$ in Metsaev's formula (\ref{Ruslan}) was found for the first time in
\cite{Metsaev:1993mj}. We are grateful to R.Metsaev for his explanations and comments.}
In the second case $(1,0)-(3/2,1/2)-(1/2,3/2)\,$,
one has $J_1 = \sum_\sigma j_\sigma^{(a)} = 3\,$, $J_2 = 2\,$,
$\min_a j^{(a)}_1=\frac{1}{2}\;$ and $\min_a j^{(a)}_2=0\;$,
so that Metsaev's formula
\begin{equation}
 2\,\max_{\sigma = 1,2} (J_{\sigma} - 2 \min_{a=1,2,3}j_\sigma^{(a)})\leqslant k \leqslant
2\min_\sigma J_\sigma \label{Ruslan}
\end{equation}
gives the solution $k=4\,$. The other two cases give no solution.
On the other hand we found the deformation
$a_1 = T^*{}^{\mu[2]}{}_\nu K_{\mu[2],\nu[2]}\,H^{\nu[3]}$ where we recall
that $H^{\nu[3]}$ is the element of $H(\gamma)$ that corresponds to
the curl of the antisymmetric gauge parameter $A_{\nu[2]}$ for the
gauge field $T$ that, in flat space, possesses two independent gauge
transformations. This candidate $a_1\,$, again, does not lead to any
nonabelian algebra since the gauge fields appear through the
curvature tensor $K_{\mu[2],\nu[2]}$. This candidate $a_1$ is
integrable and gives a vertex $a_0$ involving four derivatives --- in agreement
with the $k=4$ prediction of Formula (\ref{Ruslan}) --- and that
gives a contribution in \AdS{} corresponding to (\ref{OmegaAAAfour}), (\ref{Yu421}):
\begin{eqnarray}
 a_0 &=& K_{\mu_1\mu_2,\nu_1\nu_1} \;\left( \partial^{[\rho}T^{\mu_1\mu_2],}{}_{\sigma}\;
\partial^{[\sigma}T^{\nu_1\nu_2],}{}_{\rho} -
\partial^{[\rho}T^{\mu_1\mu_2],}{}_{\rho}\;\partial^{[\sigma}T^{\nu_1\nu_2],}{}_{\sigma} \right)
\quad,\\
\delta^{(1)}T_{\mu\nu,}{}^{\rho} &=&
K_{\mu\nu,\alpha\beta}\;\partial^{[\rho}A^{\alpha\beta]}\quad.
\end{eqnarray}
Using group theory\footnote{See e.g. the Lie program at
{\tt{http://www-math.univ-poitiers.fr/$\sim$maavl/LiE/form.html}}.},
it can be seen that this vertex is indeed nontrivial in $d=6\,$.

There is yet another vertex, this times with six derivatives: the
Born--Infeld vertex simply obtained by contracting the indices of
the three linearized curvature tensors. Using group theory again,
one can see this time  that in $d=6$ there is no way to contract the
three curvature tensors so that the result is nonvanishing. This
vertex starts being nontrivial from $d=7$ on.

Finally, there is the Lorentz minimal coupling between the hook
field and the graviton, bringing two derivatives in the Lagrangian.
Interestingly enough, this vertex appears through the following
only possible nontrivial candidate in $antigh=3$:
\begin{equation}
 a_3 ~=~ B^{*\mu}B^{\nu}\partial_{[\mu}C_{\nu]}
\end{equation}
that can be integrated via (\ref{descent3}) to give
\begin{equation}
 a_2 ~=~ C^{*\nu\mu}\left[ (A_\nu{}^{\alpha} + \frac{1}{3}\,S_\nu{}^{\alpha})\partial_{[\mu}C_{\nu]}
            + B^\alpha\partial_{[\mu}h_{\alpha]\nu}\right]\quad.
\end{equation}
When one tries to integrate $a_2$ via (\ref{descent2}), however, one
finds an obstruction to finding $a_1\,$ so that the Lorentz minimal
coupling is inconsistent in flat spacetime. As we have seen, the
Lorentz minimal couplings appears in \AdS{} (see Section
\ref{sec:simplehookST}) and are consistent when added to an
appropriate \emph{finite} tail of higher-derivative vertices, so
that the resulting coupling can be called \emph{quasi-minimal}, like
for the gravitational interactions of totally symmetric fields
\cite{Boulanger:2008tg}.

\paragraph{A remark on the nonabelianization in \AdS{} :}

We have listed above all the possible couplings between the three
types of fields $\{[2,1],[1,1]^{(1)},[1,1]^{(2)}\}$ and have seen
that, in flat spacetime, there is no nonabelian coupling among them.
The remarkable fact is that these couplings, when embedded in
\AdS{}, become related to each other and contribute to give the
nonabelian interactions we presented here in various forms. Contrary
to the totally symmetric case studied in \cite{Boulanger:2008tg}, in
the flat limit, the St\"uckelberg action can give only
\emph{abelian} interactions. The reason is that the gauge
transformations for the St\"uckelberg companion of the hook field
has got a term, in \AdS{}, that vanishes in the flat limit, being
proportional to the cosmological constant; see e.g. Eqs
(\ref{curvA})--(\ref{curvD}). Now, in the case of mixed-symmetry
fields in \AdS{} in the St\"uckelberg formulation, it is no longer
true that the linearized gauge transformations can be viewed as an
AdS covariantization of the flat space transformations. We believe
that it is responsible for that fact that, for mixed-symmetry fields
in the St\"uckelberg formulation, as opposed to the case of
totally-symmetric gauge fields, there is the possibility of scaling
away the nonabelian nature  of a vertex while at the same time retaining
a top vertex.

\section{Conclusions}
\label{sec:Conclusion}
In this paper we have obtained nonabelian gravitational interactions
for a simple mixed-symmetry gauge field in \AdS{} using various
techniques that agree with each other upon partial gauge fixing and
trivial field-redefinitions. In the St\"uckelberg formulation, the
flat limit is smooth also for the cubic action, which strengthens
the proposal of \cite{Brink:2000ag}. This is not surprising, since
the cubic vertices can smoothly be switched on and off by turning
the coupling constant, so the fact that the quadratic action allows
for a smooth flat limit implies the same for the cubic action.

Starting right away in flat space and dressing the list of all
possible cubic couplings, we indeed recuperated the
highest-derivative vertices of the \AdS{} action and found that all
the flat spacetime vertices give rise to abelian gauge algebras ---
although the gauge transformations may be linear in the gauge
fields, which appear then only through the linearized curvature like
for the Bel--Robinson or Chaplin--Manton vertices. This means that
taking the flat limit of a nonabelian action for mixed-symmetry
fields in \AdS{} trivializes the gauge algebra, in sharp contrast to
what happens for totally symmetric gauge fields in \AdS{}. This is
another instance where mixed-symmetry gauge fields in \AdS{} differ
from their totally symmetric cousins.

This is in accordance with the fact that mixed-symmetry gauge fields
in \AdS{} have only one genuine differential gauge parameter and not
several like in flat space, which makes the nonabelian coupling
problem less constrained. Generically, we expect that with one
genuine gauge parameter in the game, one can have nonabelian
interactions, like for totally symmetric fields both in flat space
and \AdS{} and for the simple mixed-symmetry field in \AdS{} studied
here. However, when one has to deal with more than one gauge
symmetry, the problem becomes too restrictive and only abelian
vertices can emerge, like for mixed-symmetry fields in flat space.

It would be interesting to make contact with the appearance of
mixed-symmetry fields within string theory through the work
\cite{Polyakov:2009pk, Sagnotti:2010at}, where their relevance was
exhibited, respectively, via vertex operators in exotic pictures and
deconstruction of tensionful string amplitudes around flat
background. Very recently, an interesting connection between string
vertex operators and higher-spin theory in \AdS{} background was
made in \cite{Polyakov:2011sm}. It would be very promising to use
this setting in order to understand better mixed-symmetry fields in
\AdS{} within string theory.

\section*{Acknowledgements}
\label{sec:Acknowledgements}
E.S. would like to thank R.Metsaev, K.Alkalaev and M.A.Vasiliev for many
valuable discussions. N.B. thanks X.Bekaert, F.Buisseret, P.P.Cook
and P.Sundell for discussions. E.S. acknowledges the Service de
M\'ecanique et Gravitation at UMONS for its hospitality. The work of
N.B. was supported in parts by an ARC contract No.
AUWB-2010-10/15-UMONS-1. the work of E.S. was supported in parts by
the RFBR grant No.11-02-00814 and the President grant No.5638. The
work of Yu.Z. was supported in parts by the RFBR grant
No.11-02-00814.

\begin{appendix}
\renewcommand{\theequation}{\Alph{section}.\arabic{equation}}

\section{Metric-like formalism for the simplest hook gauge field}
\setcounter{equation}{0}\setcounter{section}{1}\label{sec:app}

In order to make connection with the metric formulation, let us
study the Lagrangian describing the simplest hook gauge field around
\AdS{} background, namely the gauge field of Young-symmetry type
$[2,1]$, from a slightly different perspective with respect to what
is done in \cite{Brink:2000ag}. In particular, we will insist on the
role played by gauge-invariant quantities which is more parallel to
the frame-like formalism adopted in this paper. We adopt here the
manifestly symmetric convention for Young tableaux. \vspace*{.1cm}

In base-manifold component notation, the gauge transformations for
the dynamical field $\Phi_{\mu_1\mu_2,\nu}$ and the St\"uckelberg
field $\chi_{\mu_1\mu_2}$ read
\begin{eqnarray}
\delta \Phi_{\mu_1\mu_2,\nu} &=& 3\, {\mathbf{Y}}^{(2,1)}_{GL(d)}
\left[ D_{\mu_1} A_{\mu_2,\nu} +{D}_{\nu} S_{\mu_1\mu_2}+\lambda
\,g_{\mu_1\mu_2}\,\xi_{\nu} \right] \;,
\label{gtPhi}\\
\delta \chi_{\mu_1\mu_2} &=& 2\,{D}_{(\mu_1} \xi_{\mu_2)} +
2\,\lambda\, S_{\mu_1\mu_2}\;, \label{gtchi}
\end{eqnarray}
where $ {\mathbf{Y}}^{(2,1)}_{GL(d)}$ denotes the projector on the
Young tableau $(\mu_1\mu_2,\nu)$ of ${GL(d,\mathbb{R})}\,$. Next we
construct the so-called curvatures, the basic objects that are
invariant under these gauge transformations. They are
\begin{eqnarray}
K_{\mu_1\mu_2,\nu_1\nu_2,\rho} &=&
{\mathbf{Y}}^{(2,2,1)}_{GL(d)}\left[
{D}_{\rho}{D}_{\nu_2}\Phi_{\mu_1\mu_2,\nu_1} - \lambda^2 \, g_{\nu_2
\rho}\, \Phi_{\mu_1\mu_2,\nu_1} -
3\,\lambda\,g_{\mu_1\mu_2}\,{D}_{\rho}\chi_{\nu_1\nu_2} \right]
\label{Kphi}\\
K_{\mu_1\mu_2,\nu_1\nu_2} &=& {\mathbf{Y}}^{(2,2)}_{GL(d)}\left[
{D}_{\nu_2}{D}_{\nu_1}\chi_{\mu_1\mu_2} -
\frac{2}{3}\,\lambda\,{D}_{\nu_2}\Phi_{\mu_1\mu_2,\nu_1} \right]\;.
\label{Kchi}
\end{eqnarray}
{}From these curvatures we build the Einstein-like invariant tensors
\begin{eqnarray}
G_{\mu_1\mu_2,\nu} &=& 8\, K_{\mu_1\mu_2,\nu \rho,}^{\qquad\;\;\;\;
\rho} +
 12\,{\mathbf{Y}}^{(2,1)}_{GL(d)} \left[g_{\mu_1\mu_2}
K_{\nu \rho,\;\;\sigma,}^{\;\;\;\;\;\rho\,\;\;\sigma}\right] =
{\Box} \Phi_{\mu_1\mu_2,\nu} + \ldots \;,
\label{Gphi}\\
G_{\mu_1\mu_2} &=& 3\,\left[ K_{\mu_1\mu_2,\nu}^{\qquad \;\nu}
       - \frac{1}{2}\,g_{\mu_1\mu_2}\,
K_{\nu\;\;\;\,,\sigma}^{\;\;\,\nu\;\;\;\;\sigma}\right] = {\Box}
\chi_{\mu_1\mu_2} + \ldots\; \label{Gchi}
\end{eqnarray}
where $\Box$ is the covariant D'Alembertian.

These Einstein-like tensors can be seen to obey the following
identities
\begin{eqnarray}
{\nabla}^{\nu} G_{\mu_1\mu_2,\nu} -2\lambda(d-3)\,G_{\mu_1\mu_2}
&\equiv & 0\;,
\label{NoetS} \\
{\nabla}^{\nu} G_{\nu\mu ,\rho} - {\nabla}^{\nu} G_{\nu \rho,\mu}
&\equiv & 0\;,
\label{NoetA} \\
g^{\mu_1\mu_2}G_{\mu_1\mu_2,\nu} - \frac{2}{\lambda}\,(d-3)
{\nabla}^{\mu_1}G_{\mu_1 \nu} &\equiv & 0\;. \label{NoetV}
\end{eqnarray}
It is then natural to propose the following action $S[\Phi,\chi]$ (see also \cite{Brink:2000ag}):
\begin{eqnarray}
S[\Phi,\chi] = \frac{1}{2}\,\int \sqrt{-g} \left[
\Phi_{\mu_1\mu_2,\nu}\,G^{\mu_1\mu_2,\nu} +
3(d-3)\chi_{\mu_1\mu_2}\,G^{\mu_1\mu_2} \right]d^dx\,
\label{StuAct21}
\end{eqnarray}
which is invariant under the enhanced $\ads$ gauge transformations
(\ref{gtPhi}), (\ref{gtchi}) by virtue of the invariance of the
Einstein-like tensors and the identities (\ref{NoetS}),
(\ref{NoetA}), (\ref{NoetV}) which are nothing but the N\"other
identities corresponding to the parameters $S_{\mu_1\mu_2}$,
$A_{\mu,\nu}$ and $\xi_{\mu}$, respectively.

Although the Einstein-like tensors $G_{\mu_1\mu_2,\nu}$ and
$G_{\mu_1\mu_2}$ contain both fields $\Phi$ and $\chi$, it can be
seen that the Euler-Lagrange equations for the corresponding fields
simply are, as expected,
\begin{eqnarray}
0 &=& \frac{1}{\sqrt{-g}}\, \frac{\delta S[\Phi,\chi]}{\delta
\Phi^{\mu_1\mu_2,\nu}} \equiv G_{\mu_1\mu_2,\nu}\;,
\label{eomPhi}\\
0 &=& \frac{1}{\sqrt{-g}}\,\frac{\delta S[\Phi,\chi]}{\delta
\chi^{\mu_1\mu_2}} \equiv 3(d-3)G_{\mu_1\mu_2}\;. \label{eomchi}
\end{eqnarray}
By taking traces of the latter field equations and inserting the
results back in the corresponding equations, one can express the
latter as the following zero-Ricci-like equations:
\begin{eqnarray}
K_{\mu_1\mu_2,\nu\rho,}^{\qquad\;\;\;\rho} = 0 \;,\quad
 K_{\mu_1\mu_2,\nu}^{\qquad \;\nu} = 0\;.
\label{TrK}
\end{eqnarray}
As explained in \cite{Brink:2000ag}, provided the cosmological
constant is non-vanishing, the field $\chi$ is a St\"uckelberg field
that can be gauge-fixed to zero inside the action. Then, the
remaining field is $\Phi$, invariant under the gauge transformation
(\ref{gtPhi}) where only the antisymmetric parameter $A_{\mu\nu}$ is
nonzero. The field equation for the gauge-fixed action therefore is
equivalent to
\begin{equation}
0=F_{\mu_1\mu_2,\nu}\equiv 8 K_{\mu_1\mu_2,\nu\rho,}^{\qquad\;\;\;\,
\rho}\mid_{\chi=0} \quad . \label{Fronsd}
\end{equation}
%
We may use the residual gauge symmetry (under the gauge
transformation (\ref{gtPhi}) where only the antisymmetric parameter
$A_{\mu , \nu}$ is nonzero) in order to simplify the field equation
(\ref{Fronsd}). Introducing the quantity
\begin{equation}
    D_{\mu,\nu}= {D}^{\rho}\Phi_{\rho[\mu,\nu]}
                  + {D}_{[\mu}\Phi_{\nu]\rho,}^{\quad\;\rho}
\label{DeDonder}
\end{equation}
it is easy to see that it transforms like
\begin{equation}
\delta_A D_{\mu,\nu}= 2\left[ {\Box} -
(d-2)\lambda^2\right]A_{\mu,\nu}\;. \label{Mets21}
\end{equation}
The equation $\delta_A D_{\mu,\nu}=0$ is the differential equation
obtained by Metsaev for a gauge parameter $A_{\mu,\nu}$ in \AdS{}
\cite{Metsaev:1995re,Metsaev:1997nj}. Indeed, from the latter work,
we know that the differential constraint on the gauge parameter
$\vert \lambda_k\rangle$ is
\begin{eqnarray}
    \left[ {\Box} - \lambda^2 ( h^{(k)}_k - k + 1 )( h^{(k)}_k - k + d )
          +  \lambda^2 \sum_{l=1}^{\nu} h^{(k)}_l \right] \vert
\lambda_k\rangle = 0 \label{Metgp}
\end{eqnarray}
where $k$ indicates, for the Young diagram associated with the gauge
field, the maximal number of upper rows which have the same length
and $h^{(k)}_l$ ($l=1,\ldots,\nu$) are the lengths of the rows
corresponding to the Young diagram associated with the gauge
parameter $\vert \lambda_k\rangle\,$. The index $\nu$ is the integer
part of $(d-2)/2$. In the present case, we have $k=1$ since the
upper block for the gauge field has height one. The Young diagram of
the gauge parameter is obtained from the Young diagram of the gauge
field by removing one box at the end of the last row of the upper
block. We indeed get an antisymmetric, rank-2 parameter,
\textit{i.e.} $h^{1}_1=1=h^{1}_2\,$, $h^{1}_l=0$ $\forall$ $l>2$ and
the equation (\ref{Metgp}) indeed reproduces $\delta_A
D_{\mu,\nu}=0$, c.f. (\ref{Mets21}).

The field equation on the gauge field $\vert \Phi\rangle$ is, from
\cite{Metsaev:1995re,Metsaev:1997nj},
\begin{eqnarray}
    \left[ {\Box} - \lambda^2 ( h_k - k - 1 )( h_k - k -2 + d )
          +  \lambda^2 \sum_{l=1}^{\nu} h_l \right] \vert \Phi \rangle = 0
\label{Metgf}
\end{eqnarray}
where $h_l$ ($l=1,\ldots,\nu$) denote the lengths of the rows
corresponding to the gauge field  $\vert \Phi \rangle\,$. For the
example at hand, the only nonvanishing entries are $h_1=2$ and
$h_2=1\,$ with $k=1$, as we explained before. Therefore
(\ref{Metgf}) gives the equation
\begin{equation}
\left[ {\Box} + 3 \lambda^2  \right] \vert \Phi \rangle = 0\;.
\label{Metgf21}
\end{equation}
This equation can be obtained from (\ref{Fronsd}) upon gauge fixing:
Indeed, after imposing the gauge-fixing condition $D_{\mu,\nu}=0$ we
are still allowed to further fix the gauge, provided the gauge
parameter satisfies $\delta_A D_{\mu,\nu}=0\,$. With such a gauge
parameter, one can set the trace of $\vert \Phi \rangle$ to zero,
since $\delta_A
\Phi_{\mu\;\;\;,\nu}^{\;\;\;\mu}=4{D}^{\mu}A_{\mu\nu}\,$. Then, at
that stage, further gauge transformations could be performed, with a
gauge parameter still obeying $\delta_A D_{\mu,\nu}=0$ and further
satisfying ${D}^{\mu}A_{\mu,\nu}=0$.

Finally, in the gauge where $\chi_{\mu_1\mu_2}$ is vanishing and
where $\lambda\neq 0\,$, the divergence of the field equation gives
us the following constraint:
\begin{equation}
0= D^{\mu} G_{\mu\nu,\rho}|_{\chi=0} = (3-d)\lambda^2\left( D^{\mu}
\Phi_{\nu\rho,\mu} + D_{(\nu}\Phi^{'}_{\rho)}
+g_{\nu\rho}D^{\mu}\Phi^{'}_{\mu} \right)\;,\quad
\Phi^{'}_{\mu}=\Phi_{\mu\;\;\;,\nu}^{\;\;\;\mu}\;, \label{divE}
\end{equation}
which allows us to set the symmetrized divergence $D^{\rho}
\Phi_{\rho(\mu,\nu)}=-\frac{1}{2}\,D^{\rho} \Phi_{\mu\nu,\rho}$ to
zero in the gauge where $\Phi^{'}_{\mu}$ is zero. Therefore, in the
gauge ${D}_{\mu,\nu}=0\,$, $\Phi^{'}_{\mu}=0\,$, the field is
traceless and totally divergenceless.

Summarizing, the equations obtained at that point are
\begin{eqnarray}
    &D^{\rho} \Phi_{\rho\mu,\nu} = 0\,,
     \quad \Phi^{'}_{\mu} =0 \,,\quad
    {D}^{\mu}A_{\mu,\nu}=0\,, &
    \nonumber \\
    &\left( \,\Box + 3 \lambda^2  \right) \vert \Phi \rangle = 0\;,\quad
    \left[ \,\Box -  (d-2)\lambda^2\right] A_{\mu,\nu}=0\;,&
    \nonumber
\end{eqnarray}
which, as shown in \cite{Metsaev:1995re,Metsaev:1997nj}, correctly
define the unitary irreducible representation of $\ads$ associated
with the Young diagram $(2,1)\,$. We thus showed that the action
(\ref{StuAct21}), with $\lambda\neq 0$, correctly describes a
massless $(2,1)$ field propagating in $\ads$ and corresponding to a
unitary irreducible representation of the latter isometry algebra.
\vspace*{0.3cm}

Taking the flat limit $\lambda\rightarrow 0$ in (\ref{StuAct21}), we
find that the resulting action indeed describes two massless
irreducible field giving $\mathfrak{o}(d-2)$ degrees of freedom
$(2,1)\oplus (2,0)$.

\end{appendix}
\providecommand{\href}[2]{#2}\begingroup\raggedright\endgroup

\end{document}